\documentclass[journal, a4paper]{IEEEtran}
\usepackage[T1]{fontenc}
\usepackage[latin9]{inputenc}
\usepackage{color}
\usepackage{amsmath}
\usepackage{amsthm}
\usepackage{amssymb}
\usepackage{esint}
\usepackage{verbatim}
\usepackage{graphicx}
\usepackage{stfloats}
\usepackage{subfigure}
\usepackage{dsfont}
\usepackage{bm}
\usepackage{multicol}
\usepackage{multirow}
\usepackage{tabularray}
\usepackage{makecell}
\usepackage{algorithm}
\usepackage{algorithmic}
\usepackage[numbers,sort&compress]{natbib}

\makeatletter
\theoremstyle{plain}
\newtheorem{thm}{\protect\theoremname}

\theoremstyle{definition}
\newtheorem{defn}{\protect\definitionname}

\theoremstyle{remark}
\newtheorem{rem}{\protect\remarkname}

\theoremstyle{lemma}
\newtheorem{lem}{\protect\lemmaname}

\theoremstyle{assumption}
\newtheorem{assp}{\protect\assumptionname}

\theoremstyle{plain}
\newtheorem{col}{\protect\corollaryname}
\makeatother

\usepackage{babel}
\providecommand{\definitionname}{Definition}
\providecommand{\theoremname}{Theorem}
\providecommand{\remarkname}{Remark}
\providecommand{\lemmaname}{Lemma}
\providecommand{\assumptionname}{Assumption}
\providecommand{\corollaryname}{Corollary}
\newcommand\hre[1]{\textcolor{blue}{#1}}
\definecolor{sblue}{RGB}{0,0,0}

\begin{document}
%\title{\hre{Generative Semantic Communications with pre-trained Foundation Models}}
%\title{\hre{1Generative Semantic Communications with Foundation Models: Perception-Error Analysis and Resource Optimization \\2 Generative Semantic Communications with Foundation Models from a Rate-Distortion-Perception Perspective \\ 3  Perception-Error Analysis and Power Allocation for Generative Semantic Communications with Foundation Models}}
\title{Generative Semantic Communications with Foundation Models: Perception-Error Analysis and Semantic-Aware Power Allocation}
\author{Chunmei Xu,~\IEEEmembership{Member,~IEEE,}
	   Mahdi Boloursaz Mashhadi,~\IEEEmembership{Senior Member,~IEEE},
       Yi Ma,~\IEEEmembership{Senior Member,~IEEE},\\
     Rahim Tafazolli,~\IEEEmembership{Fellow,~IEEE},
     Jiangzhou Wang,~\IEEEmembership{Fellow,~IEEE}

\thanks{This work was supported by the U.K. Department for Science, Innovation, and Technology under Project TUDOR (Towards Ubiquitous 3D Open Resilient Network).}

\thanks{C.~Xu, M. Boloursaz Mashhadi, Y. Ma and, R. Tafazolli are with 5GIC \& 6GIC,
Institute for Communication Systems (ICS), University of Surrey, Guildford,
U.K. (emails:\{chunmei.xu; m.boloursazmashhadi; y.ma; r.tafazolli\}@surrey.ac.uk). %(Corresponding author: \emph{C.~Xu})
}

\thanks{J.~Wang  is with the School of Engineering, University of Kent, CT2 7NT
Canterbury, U.K. (e-mail: j.z.wang@kent.ac.uk).}

%\thanks{D.~Niyato is with the School of Computer Science and Engineering, Nanyang Technological University, Singapore 639798 (e-mail: dniyato@ntu.edu.sg).}
}

\maketitle
\thispagestyle{empty}
\begin{abstract}
Generative foundation models can revolutionize the design of semantic communication (SemCom) systems {\color{sblue}by enabling} high fidelity exchange of semantic information at ultra-low rates. In this work,  a generative SemCom framework {\color{sblue}utilizing} pre-trained foundation models is proposed, where both uncoded forward-with-error and coded discard-with-error schemes are developed for the semantic decoder.  {\color{sblue}Using the rate-distortion-perception theory, the relationship between regenerated signal quality and  transmission reliability  is characterized, which is proven to be non-decreasing.} Based on this, semantic values are  defined to {\color{sblue} quantify the semantic similarity between multimodal semantic features and the original source}. We also investigate semantic-aware power allocation problems that minimize power consumption  for ultra-low rate and high fidelity SemComs.  Two semantic-aware power allocation methods are proposed by leveraging the non-decreasing property of the perception-error relationship. {\color{sblue}Based on the Kodak dataset, perception-error functions and semantic values are obtained for image tasks.} Simulation results show that the proposed semantic-aware method significantly outperforms conventional approaches, particularly in the channel-coded case (up to 90\% power saving). %Notably/Remarkly/Appearling, the allocated power to some semantic features could be zero, suggesting that the corresponding semantic extractors can be deactivated to alleviate the communication/computation overhead. This also initially demonstrates the potential of the adaptive semantic encoding for generative SemCom systems.
\end{abstract}

\begin{IEEEkeywords}
Semantic communication, generative foundation model, rate-distortion-perception,  semantic-aware power allocation.
\end{IEEEkeywords}

%\textcolor{red}{1. Simulation part: fitted function (Fig.3 use more points? Fig.4(a) plot the fitting points? Fig. 5) 2. Intro 3. perception-theory(delete) 4. Abstract and conclusion, 5. Appendix generated data} 

\section{Introduction}

%\hre{XCM comment2: Please revise generated signal into regenerated signal }

%\hre{XCM comment3: Use "Channel-coded" and "channel-uncoded" rather than "channel coded and uncoded"}

%\hre{XCM comment4: the CLIP metric does not necessarily within the intrval of [0,1], please do not mention its range.}
%\hre{XCM: ClipScore is based on the multimodal model CLIP, which aligns images and natural language descriptions, enabling it to 'understand' images at a higher level. paper: Semantic Similarity Score for Measuring Visual Similarity at Semantic Level}

{\color{sblue}Communication systems have been developed and optimized based on Shannon information theory over the past decades, achieving remarkable success. However, the focus is primarily on the accurate reconstruction of a source signal rather than the underlying meaning of the source content. A new paradigm called semantic communication (SemCom) has emerged, shifting focus to precise content reconstruction with equivalent semantics   \cite{carnap1952outline,gunduz2022beyond}. SemCom demonstrates significant potential for achieving ultra-low compression rates and enhanced resource efficiency, maintaining effectiveness even when partial information is lost under semantic metrics. %This promising approach has garnered increasing attention from both academic  and industry.
}

{\color{sblue}Since the establishment of Shannon's information theory, researchers have pursued the development of semantic information theory through various approaches.  Initial efforts characterized semantic information by introducing semantic entropy concept using logical probability \cite{bao2011towards} and fuzzy mathematics theory \cite{de1993definition, de1974entropy}.
Theoretical advances emerged through rate-distortion theory,  examining semantic information properties via intrinsic state pairs \cite{liu2021rate} and joint probability distributions \cite{guo2022semantic}. A recent development came with Niu \emph{et al.}'s systematic framework for semantic information theory, which developed novel semantic entropy measures and comprehensive theorems in semantic source, channel, and rate-distortion coding \cite{niu2024mathematical}. Furthermore, the authors in \cite{Shao2024Theory} introduced a new conceptualization of SemCom, and formulated two fundamental problems termed language exploitation and language design, aiming to shed light on the intricate dynamics of SemCom. 
It is also recognized that semantic information should be task-dependent, corresponding to specific tasks or goals at the destination \cite{gunduz2022beyond}.  This insight motivates the integration of rate-distortion-perception theory \cite{blau2019rethinking} into SemCom, providing a theoretical framework to analyze source encoding efficiency in capturing semantic information while maintaining perceptual quality performance.
Despite these advances, a universal semantic information theory for SemCom system design remains an open challenge.}

{\color{sblue}Nevertheless, the rapid advancement of artificial intelligence (AI) has enabled significant progress in the SemCom systems, particularly through deep learning approaches. The deep learning-enabled SemCom typically employs an end-to-end architecture to jointly learn the neural network (NN)-based semantic encoder and decoder, establishing a shared knowledge base between transceivers.} The deep joint source and channel coding (JSCC) proposed in \cite{bourtsoulatze2019deep}  adopted auto-encoder NN networks for image tasks, sparking numerous deep JSCC variants for various types of sources and channel models  \cite{xie2021deep, weng2021semantic, tung2022deepwive, yang2022ofdm}. To train a deep JSCC model, the loss function was generally designed based on measurable distortion metrics such as mean square error (MSE) and peak-signal-to-noise (PSNR). {\color{sblue}While these deep JSCC approaches demonstrate superior performance compared to conventional separated source and channel coding schemes, they fundamentally adhere to Shannon's rate-distortion theory due to the use of the distortion-based loss functions.} However, distortion of SemCom systems may no longer serve as  the key performance indicator for emerging applications with specific tasks or goals, where accurate semantic information conveying becomes the primary objective.

Generative SemCom systems, utilizing deep generative AI models such as variational autoencoder (VAE), generative adversarial network (GAN), and diffusion model, {\color{sblue}show promise in preserving semantics while reducing data traffic \cite{liang2024generative}. In \cite{choi2019neural},  the authors proposed a VAE-based deep JSCC system that optimizes compression rate and error correction simultaneously, achieving robust data representations and competitive performance against conventional separated schemes. At receivers, the adopted GANs were trained using sophisticated loss functions, which combined the MSE and perceptual distances in  \cite{erdemir2023generative},   and incorporated reconstruction, synchronization and  binary discriminator errors in  \cite{tong2024multimodal}.	More recently, state-of-art diffusion models have achieved breakthrough in image \cite{rombach2022high}, audio \cite{ghosal2023text}, and video \cite{bar2024lumiere} generation tasks, offering stronger stability than GAN models in synthesizing multimedia content while preserving semantics.} A generative diffusion-guided SemCom framework was proposed, where the loss function was designed as the weighted sum of the MSE and Kullback-Leibler (KL) divergence \cite{grassucci2023generative}. This diffusion-guided SemCom was shown to achieve high robustness to extremely bad channel conditions and superior performance in generating semantically equivalent images.

However, the above deep learning-enabled SemCom systems face two challenges in their end-to-end architecture.  {\color{sblue}Firstly, analog modulations are required for gradient computation and back-propagation during training, which conflicts with modern digital communication systems. Secondly, training semantic encoders and decoders for fading and noisy channels demands substantial computational resources while showing poor generalization across different data sources and channel models.}  To address these, a potential solution is to adopt foundation models like bidirectional encoder representations from transformers (BERT) and generative pre-trained transformer (GPT). These models{\color{sblue}, trained on vast amounts of diverse datasets, can capture general patterns and thereby enable knowledge base sharing between transceivers.} Particularly promising are the generative foundation models based on diffusion models, such as  DALL$\cdot$E and Sora, which can synthesize high perceptual quality signals by exchanging extremely compressed textual prompts. These advances motivate to design SemCom systems using foundation models,  conveying semantic information with minimal data traffic.

In this work, we propose a generative SemCom framework that uses  pre-trained foundation models for semantic encoding and decoding. {\color{sblue}With  semantic encoder and decoder fixed, transmission reliability emerges as the key factor affecting the perceptual quality of regenerated signals. We analyze their   relationship through rate-distortion-perception theory and develop semantic-aware resource allocation strategies to optimize power consumption while maintaining semantic performance. For the ultra-low rate and high fidelity SemCom, we focus on high-reliability schemes: uncoded binary phase shift keying (BPSK) for channel-uncoded case and finite block length coding \cite{polyanskiy2010channel}  for channel-coded cases respectively.  The main contributions of this work are summarized as follows: }

%Given the semantic encoder and decoder, the transmission scheme (including the channel coding/decoding, modulation/demodulation modules) and wireless channels, retain its influence on the perceptual quality of the regenerated signal. In other words, the transmission reliability becomes the only impact factor influencing the perceptual quality, the theoretical analysis of which has not yet been investigated. To fill this gap, we provide the theoretical analysis to model their mathematical relationship based on the rate-distortion-perception theory, and characterize the semantic value of semantic features to measure the semantic information accordingly. We investigate the semantic-aware resource allocation problems aiming at minimizing the total power consumption while guaranteeing the semantic performance of the generated signal considering both channel-uncoded and channel-coded cases. Given that semantic data streams are transmitted at ultra-row rates in the proposed generative SemCom framework, the transmission schemes with high reliability are considered. Specifically, the uncoded binary phase shift keying (BPSK)  and finite block length coding \cite{polyanskiy2010channel} are considered under channel-uncoded and channel-coded cases respectively. The contributions of this work are summarized as follows: 

\begin{itemize}
    \item  {\color{sblue}A generative SemCom framework is proposed using pre-trained foundation models as semantic encoder and decoder, leveraging their shared knowledge bases and generalization capabilities without additional training.}  Both uncoded forward-with-error and coded
    discard-with-error schemes are developed  in the semantic decoder.
     
    \item {\color{sblue}The relationship between transmission reliability and regenerated signal quality is analyzed through rate-distortion-perception theory, proving} that perception value deteriorates with transmission errors. Semantic values of transmitted
    and received data streams are defined based on the perception value to quantify their semantic similarities with the original source.
    
    \item {\color{sblue}The semantic-aware power allocation problems under both channel-uncoded and channel-coded cases are investigated to minimize total power consumption while maintaining the semantic performance.} Two methods are developed by leveraging the non-decreasing property of the perception-error relationship. 
     
    \item Perception-error functions and semantic values  under both uncoded forward-with-error and coded discard-with-error schemes are obtained through simulations on the Kodak dataset. Compared to conventional approaches, the proposed semantic-aware bisection method achieves power savings of up to 10\% and 90\% in channel-uncoded and channel-coded cases, respectively, 
    \end{itemize}

\section{Generative SemCom Framework}\label{sec:system_model}

\begin{figure*}[tbp]
\centering
\includegraphics[width=0.9\textwidth]{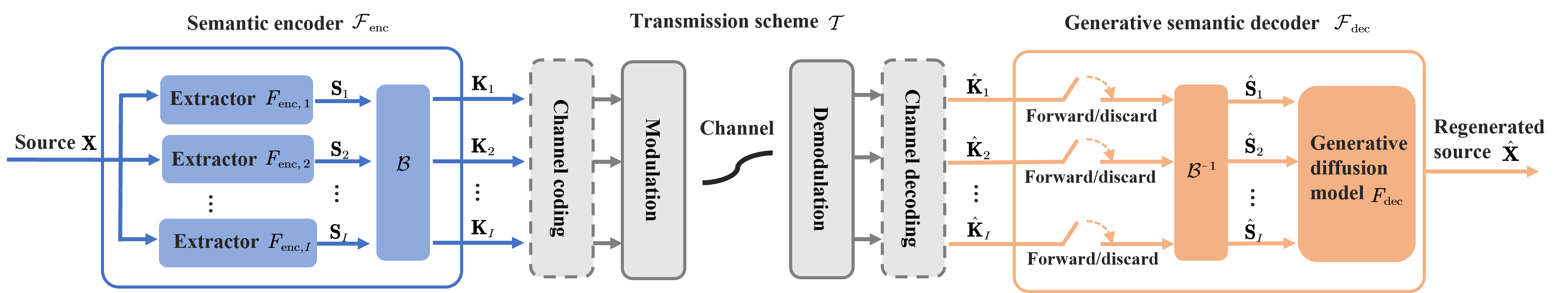}
\caption{{\color{sblue}The proposed generative semantic communication framework with pre-trained foundation models.}}
\label{fig:system_model}
\end{figure*}

This section introduces the proposed generative  SemCom framework as depicted in Fig. \ref{fig:system_model}, which consists of semantic encoder, transmission scheme, and semantic decoder. %The pre-trained foundation models are customized in both semantic encoder and decoder, and the transmission scheme includes conventional modules such as channel coding and decoding as in conventional communication systems. %which consists of the AI-based semantic encoder, wireless transmission and the diffusion model based semantic decoder. Note that the AI models adopted in the semantic encoder and decoder are pre-trained without any further data training. 

\subsection{Semantic Encoder}
{\color{sblue}The semantic encoder employs $I$ semantic extractors to} extract semantic features from the inputted source signal $\mathbf{X}$ using the pre-trained foundation models $F_{\mathrm{enc},i}$. The $i$-th semantic feature can be expressed by 
\begin{equation}
\mathbf{S}_{i}=F_{\mathrm{enc},i}(\mathbf{X}\mid \bm{\theta}_i^*),\end{equation}where $\bm{\theta}_i^*$ is the NN parameters of the $i$-th foundation model. {\color{sblue}For image signals, these semantic features may include prompts, edge maps and} segmented semantics, extracted using various pre-trained models such as image-to-text transformers \cite{li2022blip}, \cite{weissman2023textual}, the Holistically-nested Edge Detection (HED) model \cite{xie2015holistically} and the DeepLab model \cite{chen2017deeplab}, respectively. {\color{sblue}To ensure compatibility with existing digital communication systems,} each semantic feature $\mathbf{S}_i$
is converted into a bit sequence, termed a semantic data stream $\mathbf{K}_{i}$,  {\color{sblue}through the operation }
$\mathbf{K}_{i}=\mathrm{\mathcal{B}}(\mathbf{S}_{i})$, where $\mathrm{\mathcal{B}}(\cdot)$
is the binary mapping function such as ASCII, Unicode, and quantization. % if the data type of $\mathbf S_{i}$ is text, it can be converted into the bit streams by the ASCII or Unicode encoding schemes. If the output is a real vector, the quantization can be applied to obtain the bit streams. 

Each semantic data stream contributes differently to the perceptual quality of the regenerated signal {\color{sblue}when evaluated} under a specific semantic metric, which is highly related to the inference goal or task at the destination. This makes it fundamentally different from conventional communication systems {\color{sblue}that equally treat all data streams. To quantify the contribution of each data stream, we introduce  the semantic value  $L_i$ to characterize its semantic similarity with the original source signal (defined in the Sec. \ref{sec3-B}). A larger $L_{i}$ indicates that the $i$-th data stream} has a greater impact on the perpetual quality of the regenerated signal, implying that it is more
important. %\hre{The semantic extractors are either activated or deactivated based on the perceptual requirement or the channel conditions, indicating that semantic encoding rate is adaptive. Intuitively, the more important semantic data streams are more likely to be extracted and transmitted especially when the channel conditions are bad. This can not only save the wireless resource but also reduce the computational overhead at the transmitter. } 
\vspace{-5pt}
\subsection{Transmission Scheme} 

{\color{sblue}In the proposed generative SemCom framework, the multi-stream transmission is modeled as:}

\begin{equation}
[\hat{\mathbf{K}}_{1},\hat{\mathbf{K}}_{2},\dots,\hat{\mathbf{K}}_{I}]=\mathcal{T}([\mathbf{K}_{1},\mathbf{K}_{2},\dots,\mathbf{K}_{I}]),
\end{equation}
where $\mathcal{T}(\cdot)$ is the transmission scheme {\color{sblue} that maps} from the transmitted data streams to the received ones, {\color{sblue} which encompasses} channel coding,  modulation,  demodulation, and channel decoding. These data streams are transmitted orthogonally to eliminate the inter-stream interference. 

Due to the fading and noisy effects of  the wireless channels, the received semantic data stream $\hat{\mathbf{K}}_i$ may contain errors.  The probability of receiving $\hat{\mathbf{K}}_{i}$ is denoted as 
$\mathbb P (\hat{\mathbf{K}}_{i}\mid\mathbf{K}_{i};\mathcal T)$, which depends on the bit error rate (BER) $\psi_{i}$ {\color{sblue}($\le 0.5$)} in the channel-uncoded case.  The block error rate (BLER) of the $i$-th semantic data stream is denoted as $\Psi_i=\mathbb P (\hat{\mathbf{K}}_{i}\neq \mathbf{K}_{i} ; \mathcal T)$. {\color{sblue}While hybrid automatic repeat request (HARQ) mechanisms can ensure correct transmission, they introduce additional latency. Alternatively, the transmission reliability can be improved by adopting adaptive coding and modulation schemes based on channel conditions.}

\subsection{{\color{sblue}Generative} Semantic Decoder}

{\color{sblue}At the generative semantic decoder, signal regeneration is performed using the received data streams $\hat{\mathbf{K}}_i$. To handle transmission errors, we propose two distinct processing schemes:  uncoded forward-with-error scheme for channel-uncoded case and  coded discarded-with-error scheme  for channel-coded case. This differentiation arises because transmission errors in channel-uncoded cases cannot be detected, while channel-coded cases experience burst errors due to codeword correlation. These burst errors may even degrade performance (demonstrated in Fig. \ref{fig:comparison_discard} in Sec. \ref{sec:VI}).}

For uncoded forward-with-error scheme,
all received semantic data streams $\hat{\mathbf{K}}_{i}$ are first reconverted into semantic features $\hat{\mathbf{S}}_i=\mathcal{B}^{-1}(\hat{\mathbf{K}}_{i})$, where $\mathrm{\mathcal{B}^{-1}}\left(\cdot\right)$ is the inverse operation of $\mathrm{\mathcal{B}}\left(\cdot\right)$. These features
are then forwarded to the generative foundation model $F_{\mathrm{dec}}$ to synthesize the generated signal $\hat{\mathbf{X}}$, expressed as
\begin{equation}
\hat{\mathbf{X}}=F_{\mathrm{dec}}(\hat{\mathbf{S}}_{1},\hat{\mathbf{S}}_{2},\dots,\hat{\mathbf{S}}_{I}; \bm{\omega}^*)\triangleq 
\mathcal F_{\mathrm{dec}}(\hat{\mathbf{K}}_{\mathcal I}),
\end{equation}where $\bm{\omega}^*$ represents the NN parameters of the generative model, and $\hat{\mathbf{K}}_{\mathcal I}\triangleq\{\hat{\mathbf{K}}_i,i\in\mathcal I\}$ is the concatenated received data streams.  {\color{sblue}For  the coded discard-with-error scheme,  semantic data streams containing errors are discarded to mitigate the impact of burst errors.} Letting $\mathcal I_c$ denote the index set of error-free received semantic data streams, the regenerated signal $\hat{\mathbf{X}}$ is expressed as 
\begin{equation}
\hat{\mathbf{X}}=F_{\mathrm{dec}}(\left\{\mathbf{S}_j\right\}_{j\in\mathcal I_c}; \bm{\omega}^*)\triangleq 
\mathcal F_{\mathrm{dec}}(\mathbf{K}_{\mathcal I_c}),
\end{equation}where  $\mathbf{K}_{\mathcal I_d}\triangleq\{\mathbf{K}_j,j\in\mathcal I_c \}$. The semantic decoder does not synthesize any signal when $\mathcal I_c=\emptyset$.

Within the proposed generative SemCom framework, the semantic information of the semantic data streams is lossy due to transmission errors. This implies that the semantic values of received data streams are reduced, i.e., $\hat{L}_{i,\mathrm{forward}}\leq L_i$ and $\hat{L}_{i,\mathrm{discard}}\leq L_i$. Here, $\hat{L}_{i,\mathrm{forward}}$ and $\hat{L}_{i,\mathrm{discard}}$ are the semantic values of the $i$-th semantic data stream $\hat{\mathbf{K}}_i$ under uncoded forward-with-error and coded discard-with-error schemes, respectively. %The definitions of the semantic values are given in the next section.
\section{Perception-Error Analysis and Semantic Value}\label{sec:semantic_tradeoff}

This section characterizes how transmission errors affect the perceptual quality of the regenerated signal through  rate-distortion-perception theory.   Based on the analyzed perception-error relationship, the semantic values of the transmitted and received semantic data streams are defined to quantify their semantic similarity with the original source. 
\subsection{{\color{sblue}Rate-Distortion-Perception Function}}
{\color{sblue}The rate-distortion-perception theory \cite{blau2019rethinking}\cite{Chen2022rate}, an extension of Shannon's rate-distortion theory, incorporates perceptual distance as an additional constraint. This theory examines three key metrics between the source signal $\mathbf{X}$ and constructed signal $\hat{\mathbf{X}}$: rate (quantified by mutual information), distortion (measured by distortion distance), and perception (evaluated by perceptual distance). The rate-distortion-perception trade-off, which  minimizes mutual information subject to distortion and perception constraints, is formulated as:

\begin{subequations}\label{eq:R_D_P_function}
    \begin{align}
      R  (D, P) \triangleq \min_{P_{\hat{\mathbf{X}}\mid\mathbf{X}}} & \,\,I(\mathbf{X}; \hat{\mathbf{X}})\\
     \mathrm{s.t.}\,\, &\mathbb{E}[d(\mathbf{X}, \hat{\mathbf{X}})]\leq D
      \\
      &%\mathbb{E}_{\color{red}{p\left(\hat{\mathbf{X}}\left(\hat{\mathbf{K}}_i\right)\mid\mathbf{X}\right)}}\left[\delta\left({\mathbf{X}}, {\hat{\mathbf{X}}}\right)\right]\leq P, 
      \delta(P_{\mathbf{X}}, P_{\hat{\mathbf{X}}})\leq P, 
    \end{align}
\end{subequations}where $P_{\mathbf{X}}$ and $P_{\hat{\mathbf{X}}}$ denote the distributions of the source signal $\mathbf{X}$ and regenerated signal $\hat{\mathbf{X}}$ respectively, and $P_{\hat{\mathbf{X}}\mid\mathbf{X}}$ represents their conditional distribution. The measurable distortion function $d(\mathbf{X}, \hat{\mathbf{X}})$, typically designed as the squared error, is constrained by $D$.

$\delta(P_{\mathbf{X}}, P_{\hat{\mathbf{X}}})$ represents the perceptual distance constrained by $P$. It can be implemented  using the distribution-based  metrics such as KL divergence and Wasserstein distance \cite{blau2019rethinking}\cite{Chen2022rate}. As an alternative, $\delta(P_{\mathbf{X}}, P_{\hat{\mathbf{X}}})$ can be designed as  non-distribution-based perceptual distances using the contrastive language-image pre-training (CLIP) similarity \cite{wang2023exploring} or multi-scale
structural similarity (MS-SSIM) \cite{wang2003multiscale}, which are expressed as:
	\begin{equation}
		\delta(P_{\mathbf{X}}, P_{\hat{\mathbf{X}}})\triangleq 1-\mathbb E[ \mathrm{CLIP}(\mathbf{X},\hat{\mathbf{X}})],
	\end{equation}
	\begin{equation}
		\delta(P_{\mathbf{X}}, P_{\hat{\mathbf{X}}})\triangleq 1 - \mathbb E[ \mathrm{MS\text{-}SSIM}(\mathbf{X},\hat{\mathbf{X}})],
	\end{equation}
where $\mathrm{CLIP}(\mathbf{X},\hat{\mathbf{X}})$ and $\mathrm{MS\text{-}SSIM}(\mathbf{X},\hat{\mathbf{X}})$ represent the CLIP and MS-SSIM similarity functions. 

The CLIP similarity is obtained using the CLIP NN model that jointly trains image and text encoders to embed both modalities into a shared, high-dimensional feature space where semantically similar content clusters together  \cite{radford2021learning}. Using contrastive learning on paired image-text data, it learns to maximize cosine similarity between matching pairs while minimizing it for non-matching pairs. Denoting the learned model $F_\mathrm{clip}$   with NN parameters $\bm{\theta}_\mathrm{clip}$, the CLIP similarity function can be expressed as:}
{\color{sblue}\begin{equation}\label{eq:clip}
\mathrm{CLIP}(\mathbf{X},\hat{\mathbf{X}})=\frac{F_\mathrm{clip}(\mathbf{X};\bm{\theta}_\mathrm{clip})\cdot F_\mathrm{clip}(\hat{\mathbf{X}};\bm{\theta}_\mathrm{clip})}{\Vert F_\mathrm{clip}(\mathbf{X};\bm{\theta}_\mathrm{clip})\Vert \Vert F_\mathrm{clip}(\hat{\mathbf{X}};\bm{\theta}_\mathrm{clip})\Vert},
\end{equation}where $\Vert\cdot\Vert$ is the $\ell_2$ norm. The resulting score ranges from -1 to 1, with higher values indicating stronger semantic similarity.  In addition, the MS-SSIM is widely used for image quality assessment in image processing and computer vision. $\mathrm{MS\text{-}SSIM}(\mathbf{X},\hat{\mathbf{X}})$ evaluates the perceptual similarity between the source signal $\mathbf{X}$ and the reconstructed signal $\hat{\mathbf{X}}$  by analyzing luminance, contrast and structure across multiple scales. The final MS-SSIM score combines these components using weighted products, yielding a value between 0 and 1, where 1 indicates perfect perceptual similarity and 0 indicates no structural similarity. Note that when no signal is regenerated (denoted as $\hat{\mathbf{X}}_{\emptyset}$), the CLIP and MS-SSIM similarities satisfy $\mathrm{CLIP}(\mathbf{X},\hat{\mathbf{X}}_{\emptyset})=0$ and $\mathrm{MS\text{-}SSIM}(\mathbf{X},\hat{\mathbf{X}}_{\emptyset})=0$.}
\begin{comment}
\begin{equation}\label{eq:ms-ssim}
\mathrm{MS\text{-}SSIM}(\mathbf{X},\hat{\mathbf{X}})=\left[l_M(\mathbf{X},\hat{\mathbf{X}}) \right]^{\alpha_M} 
\prod_{j=1}^{M} \left[ c_j(\mathbf{X},\hat{\mathbf{X}}) \right]^{\beta_j}\left[ s_j(\mathbf{X},\hat{\mathbf{X}}) \right]^{\gamma_j},
\end{equation}where $F_\mathrm{clip}(\cdot)$ represents a pre-trained model on a large text-image dataset that encodes images or prompts into feature representations \cite{radford2021learning}. When all received data streams are discarded such that none signal termed as $\hat{\mathbf{X}}_{\emptyset}$ is regenerated, the CLIP value becomes $\mathrm{CLIP}(\mathbf{X},\hat{\mathbf{X}}_{\emptyset})=1$.
\end{comment}

\subsection{{\color{sblue}Perception-Error Analysis}}
	
SemComs aim to convey the semantic meanings with {\color{sblue}minimal} semantic loss, i.e., the smallest perceptual distance,  regardless of distortion. {\color{sblue}Since distortion is not constrained in this context, we can set $D=\infty$ and reformulate the rate-distortion-perception trade-off in \eqref{eq:R_D_P_function} as:}
\begin{subequations}\label{eq:P_R_D_function}
    \begin{align}
      %P & \left(R, D | \mathrm{Task, Interest}\right) \nonumber\\ &\triangleq \min_{\color{red}{p\left(\hat{\mathbf{X}},\mathbf{X}\right)}} \mathbb{E}\left[\delta\left(\mathbf{X}, \hat{\mathbf{X}}|{\mathrm{Task, Interest}}\right)\right] \\
      P  (R)\triangleq \min_{P_{\hat{\mathbf{X}}\mid\mathbf{X}}}  &\,\,{\color{sblue}\delta(P_{\mathbf{X}}, P_{\hat{\mathbf{X}}})} \\
     \mathrm{s.t.}\,\, &I(\mathbf{X}; \hat{\mathbf{X}})\leq R, \label{eq:P_R_D_cons2}
    \end{align}
    \end{subequations}{\color{sblue}which aims}  to find the conditional distribution $P_{\hat{\mathbf{X}}\mid \mathbf{X}}$ under the rate constraint $R$. In the proposed generative SemCom framework, $P_{\hat{\mathbf{X}}\mid \mathbf{X}}$ is jointly determined by the semantic encoder, transmission scheme, semantic decoder, and the channel conditions. Since the semantic encoder $\mathcal F_{\mathrm{enc}}$ and generative semantic decoder $\mathcal F_{\mathrm{dec}}$ are designed using pre-trained foundation models, only the transmission scheme $\mathcal T$ and channel influence the regenerated signal's perceptual quality. This means that $P_{\hat{\mathbf{X}}\mid\mathbf{X}}$ depends solely on the conditional distribution $P_{\hat{\mathbf{K}}_{\mathcal I}\mid \mathbf{K}_{\mathcal I}}$  
     {\color{sblue} under the proposed generative SemCom framework}.

The inputted source signal, transmitted data streams, received data streams, and generated signal form a Markov chain such that $\mathbf{X}\rightarrow\mathbf{K}_\mathcal{I}\rightarrow \hat{\mathbf{K}}_\mathcal{I} \rightarrow \hat{\mathbf{X}}.
$
{\color{sblue}Applying the data-processing inequality \cite{thomas2006elements} yields two inequalities:}
$I(\mathbf X; \hat{\mathbf{X}}) \le I(\mathbf X; \hat{\mathbf{K}}_{\mathcal I})$ and $I(\mathbf X; \hat{\mathbf{K}}_{\mathcal I}) \le I(\mathbf K; \hat{\mathbf{K}}_{\mathcal I})$. {\color{sblue}These two inequalities combine to give:}
\begin{equation}\label{eq:I_inequality}
I(\mathbf X; \hat{\mathbf{X}}) \le I(\mathbf K_\mathcal{I}; \hat{\mathbf{K}}_\mathcal{I}).
\end{equation}
{\color{sblue}The right-hand side of (\ref{eq:I_inequality}) further satisfies:  
\begin{equation}\label{eq:I_inequality1}
I(\mathbf K_\mathcal{I}; \hat{\mathbf{K}}_\mathcal{I}) \le \sum_{i\in\mathcal I} I(\mathbf K_i; \hat{\mathbf{K}}_i),
\end{equation}}where the equality holds if and only if the semantic data streams are independent. {\color{sblue}Combining \eqref{eq:I_inequality} and \eqref{eq:I_inequality1} yields: 
\begin{equation}\label{eq:11}
	I(\mathbf X; \hat{\mathbf{X}})  \le \sum_{i\in\mathcal I} I(\mathbf K_i; \hat{\mathbf{K}}_i).
	\end{equation}}

By replacing \eqref{eq:P_R_D_cons2} with \eqref{eq:11}, problem (\ref{eq:P_R_D_function}) can then be approximated by: 
\begin{subequations}\label{eq:P_D_function}
    \begin{align}
      P (R)\triangleq  \min_{P_{\hat{\mathbf{K}}_{\mathcal I}\mid\mathbf{K}_{\mathcal I}}} & {\color{sblue}\delta(P_{\mathbf{X}}, P_{\hat{\mathbf{X}}})} \\
     \mathrm{s.t.}\,\, & \sum_{i\in\mathcal I} I(\mathbf K_{i}; \hat{\mathbf{K}}_{i})\le R, \label{eq:P_D_cons1}
\end{align}\end{subequations}where $P_{\hat{\mathbf{K}}_{\mathcal I}\mid \mathbf{K}_{\mathcal I}}=\prod_{i\in\mathcal I}P_{\hat{\mathbf{K}}_{i}\mid \mathbf{K}_{i}}$   {\color{sblue}due to the independent transmission of all data streams.  The objective is to find the optimal conditional distribution $P_{\hat{\mathbf{K}}_{\mathcal I}\mid \mathbf{K}_{\mathcal I}}$ that minimizes the perceptual distance, termed as the perception value, subject to the rate constraint $R$.}  
\begin{lem}\label{lem1}
	The perception-rate function $P(R)$ is  {\color{sblue}non-increasing} with rate $R$.
	\begin{proof}
		The perception-rate function $P(R)$ represents the minimum perceptual distance achievable over a feasible set of conditional distributions $P_{\hat{\mathbf{K}}_{\mathcal I}\mid \mathbf{K}_{\mathcal I}}$. {\color{sblue}As $R$ increases, this feasible set monotonically expands, and consequently, $P(R)$ is a non-increasing function of $R$}.
	\end{proof}
\end{lem}

The conditional distribution $P_{\hat{\mathbf{K}}_{\mathcal I}\mid \mathbf{K}_{\mathcal I}}$ is determined by both transmission schemes and channel conditions, naturally motivating the study of adaptive SemComs with channel feedback. However, simply combining  conventional adaptive techniques with the generative SemCom might not offer additional semantic performance gains. This is because semantic features may exhibit varying levels of importance. Conventional adaptive techniques without considering semantic importance can assign less-important features to good channel conditions, resulting in inefficient uses of radio resources. This interesting problem invokes a new research direction in the scope of generative SemCom. This work is, however, focused on the generative SemCom with fixed  coding rate, and investigate how transmission reliability affects the perceptual quality of the regenerated signal. To establish their relationship, we introduce the following assumption.
\begin{assp}\label{assp2} 
	{\color{sblue}The bits within transmitted semantic data streams are assumed to be independent. The $j$-th bit of the $i$-th semantic data stream, denoted as $\mathbf{K}_{ij}$, follows a Bernoulli distribution with probability $\phi_{ij}$ of being $1$ and  $1-\phi_{ij}$ of being $0$. The probability $\Phi_i$ of sequence $\mathbf{K}_i$ is given by $\Phi_i=\prod_{j=1}^{K_i} (\mathbf{K}_{ij}\phi_{ij}+ (1-\mathbf{K}_{ij})(1-\phi_{ij}))$.
	}
\end{assp}

\begin{lem}\label{lem2} Under Assumption \ref{assp2}, The mutual information between $\mathbf{K}_{i}$ and $\hat{\mathbf{K}}_{i}$ for both the uncoded forward-with-error and coded discard-with-error schemes are  given by: 
	{\color{sblue}\begin{equation} \label{eq:13}   
	I(\mathbf{K}_{i};\hat{\mathbf{K}}_{i})=
		\sum_{j=1}^{K_{i}}H(\phi_{ij})-H(\psi_{ij}),
	\end{equation}
and 
\begin{equation} \label{eq:14}   
	I(\mathbf{K}_{i};\hat{\mathbf{K}}_{i})= 
		H(\Phi_{i})-\Psi_{i}H(\Phi_{i}),  
\end{equation}respectively,} where $H(\cdot)$ is the entropy function. The mutual information $I(\mathbf{K}_{i};\hat{\mathbf{K}}_{i})$  decreases with increasing BER  $\psi_{ij}$ or BLER $\Psi_{i}$. The proof is provided in Appendix \ref{Proof_Lemma2}.
\end{lem}
\begin{comment}
	\begin{equation} \label{eq:13}   
		I\left(\mathbf{K}_{i};\hat{\mathbf{K}}_{i}\right)=\begin{cases}
			\sum_{j=1}^{K_{i}}H\left(\phi_{ij}\right)-H\left(\psi_{ij}\right)& \textrm{\small forward-with-error}\\
			H\left(\Phi_{i}\right)-\Psi_{i}H\left(\Phi_{i}\right) & \textrm{\small discard-with-error}
	\end{cases}\end{equation}

Substituting \eqref{eq:13} and \eqref{eq:14} back to the perception-rate function \eqref{eq:P_D_function} yields the perception-error functions for both the uncoded forward-with-error and coded discard-with-error scheme, which are respectively given by:
	\begin{subequations}\label{eq:P_E_function}
		\begin{align}
			P \left(R\right)\triangleq  \min_{P_{\hat{\mathbf{K}}_{\mathcal I}\mid\mathbf{K}_{\mathcal I}}} & {\color{sblue}\delta\left(P_{\mathbf{X}}, P_{\hat{\mathbf{X}}}\right)} \\
			\mathrm{s.t.}\,\, & \sum_{i\in\mathcal I} \sum_{j=1}^{K_{i}}H\left(\phi_{ij}\right)-H\left(\psi_{ij}\right)\le R, 
	\end{align}\end{subequations}
and 
	\begin{subequations}\label{eq:P_E_function}
		\begin{align}
			P \left(R\right)\triangleq  \min_{P_{\hat{\mathbf{K}}_{\mathcal I}\mid\mathbf{K}_{\mathcal I}}} & {\color{sblue}\delta\left(P_{\mathbf{X}}, P_{\hat{\mathbf{X}}}\right)} \\
			\mathrm{s.t.}\,\, & \sum_{i\in\mathcal I} \sum_{j=1}^{K_{i}}H\left(\phi_{ij}\right)-H\left(\psi_{ij}\right)\le R, 
	\end{align}\end{subequations}
\end{comment}	
 It is challenging to obtain the optimal distribution due to its dependence on source distribution $P_{\mathbf X}$ and the implicit mapping of pre-trained foundation models $F_{\mathrm{enc},i}$ and $F_{\mathrm{dec}}$. {\color{sblue}According to Assumption \ref{assp2} and Lemma \ref{lem2},  finding the optimal solution for $P_{\hat{\mathbf{K}}\mid \mathbf{K}}$ becomes equivalent to optimizing the transmission scheme $\mathcal T$ based on the channel conditions to obtain the optimal BER $\psi_{ij}$ or BLER $\Psi_{i}$.}  For any {rate $R$}, there exists a corresponding optimal distribution solution {\color{sblue}$P_{\hat{\mathbf{K}}_{\mathcal I}| \mathbf{K}_{\mathcal I}}$} (or equivalent optimal BER ${\color{sblue}\psi_{ij}}$ and BLER {\color{sblue}$\Psi_{i}$}).  {\color{sblue}Consequently,  we can reframe the perception-rate function as a perception-error function to better characterize how transmission reliability affects the perceptual quality of the regenerated signal.} Denote the perception-error function as {\color{sblue}$P_{\mathrm{forward}}(\{\psi_{ij} \}_{i,j}| R)$}  for uncoded forward-with-error scheme and {\color{sblue}$P_{\mathrm{discard}}(\{\Psi_{i}\}_i| R)$}  for coded discard-with-error scheme. Based on  {Lemma \ref{lem1} and {Lemma \ref{lem2}},  the following corollary is established.

\begin{col}\label{thm:decreasing}
     {\color{sblue}$P_{\mathrm{forward}}(\{\psi_{ij}\}_{i,j} | R)$ and $P_{\mathrm{discard}} (\{\Psi_{i}\}_i | R )$ are non-decreasing with $\psi_{ij}$ and $\Psi_i$,} respectively. This monotonic relationship demonstrates that the perceptual quality of the regenerated signal deteriorates as transmission errors increase.  
    %The perception values $P_{\mathrm{forward}}\left(\left\{\psi_{i,j}\right\}_{i,j}\right)$ and $P_{\mathrm{discard}}\left(\left\{\Psi_{i}\right\}_i\right)$ are non-decreasing in $\psi^*_{i,j}$ and $\Psi^*_i$    under forward-with-error and discard-with-error schemes, respectively.
\end{col}

\subsection{Semantic Value \label{sec3-B}}
{\color{sblue}%To quantify the effectiveness of extracted semantic features in representing the source signal, we evaluate their semantic similarities  with the original source. 
	Based on the previously analyzed perception-error functions, we define semantic values for both transmitted and received semantic data streams to quantify their  semantic similarities with the original source signal.}
\begin{defn}\label{den:perception value}
The semantic value of the $i$-th transmitted semantic data stream $\mathbf{K}_i$ is defined as:
\begin{equation}
    L_i = 1-{\color{sblue}{P}^{(i)}},
\end{equation}where ${\color{sblue}{P}^{(i)}=\delta(P_{\mathbf{X}}, P_{{\mathbf{X}}^{(i)}})}$ is the perception value of regenerated signal ${\mathbf{X}}^{(i)}=\mathcal F_{\mathrm{dec}}\left(\mathbf{K}_i\right)$ synthesized using the $i$-th semantic data stream $\mathbf{K}_i$. 
\end{defn}

\begin{defn}
    The semantic values of the $i$-th received semantic data stream $\hat{\mathbf{K}}_i$ under the uncoded forward-with-error and  coded discard-with-error schemes are defined  as: 
\begin{equation}
\hat{L}_{i,\mathrm{forward}}(\{{\color{sblue}\psi_{ij}}\}_{j})= 1-{\color{sblue}P^{(i)}_{\mathrm{forward}}(\{\psi_{ij}\}_{j})}, 
\end{equation}
and
\begin{equation}
\begin{aligned} \hat{L}_{i,\mathrm{discard}}({\color{sblue}\Psi_{i}})&=
1-{\color{sblue}P^{(i)}_{\mathrm{discard}}\left(\Psi_{i}\right)},
\end{aligned}\end{equation}where ${\color{sblue}P^{(i)}_{\mathrm{forward}}(\left\{\psi_{ij}\right\}_{j})=\delta(P_{\mathbf{X}}, P_{\hat{\mathbf{X}}^{(i)}})}$ is the perception value of the regenerated signal $\hat{\mathbf{X}}^{(i)}=\mathcal F_{\mathrm{dec}}(\hat{\mathbf{K}}_i)$ synthesized using the $i$-th received semantic data stream $\hat{\mathbf{K}}_i$ with BER $\psi_{ij}$. {\color{sblue}$P^{(i)}_{\mathrm{discard}}(\Psi_{i})=\Psi_{i}\delta(P_{\mathbf{X}}, P_{\hat{\mathbf{X}}_{\emptyset}})+(1-\Psi_{i})\delta(P_{\mathbf{X}}, P_{{\mathbf{X}}^{(i)}})$ where $\delta(P_{\mathbf{X}},P_{\hat{\mathbf{X}}_{\emptyset}})=1$.}
\end{defn}

\begin{rem}\label{rem1}
The semantic value of the received semantic data stream is non-increasing in {\color{sblue}$\psi_{ij}$ or $\Psi_i$, indicating that transmission errors result in loss of semantic information.}
\end{rem}
\begin{rem}\label{rem2}	
The semantic values differ among the semantic data streams,  {\color{sblue}reflecting their varying levels of importance in signal regeneration.}   
\end{rem}
\begin{rem}
	For any given semantic data stream, its semantic value varies across different perceptual measurements, implying {\color{sblue}that its importance is related to the specific tasks or goals at the receiver.}
	
	%The semantic values depend on the , which are highly related to the goals or tasks at the destination. , which has different semantic values under different perceptual measurements, 
\end{rem}

\section{Problem Formulations}
\begin{comment}
\textcolor{blue}{ Based on the above observations, we give \textcolor{blue}{three definitions of semantic information in terms of bit, modulated symbol and the packet} in the following. 
\begin{defn}
Assuming that the semantic information is independent among the bits with equal semantic units, the semantic
information $\alpha_{i}$ of the $i$th semantic feature is modeled
by the amount of semantic units per bit , i.e., $\alpha_{i}\triangleq L_{i}/K_{i}$. 
\end{defn}
%
\begin{defn}
Assuming that the semantic is independent among the modulated symbols with equal semantic units,
the semantic information of the $i$th transmitted and received are defined by the amount of semantic units per symbol,
i.e., $\alpha_{i}\triangleq L_{i}\log_{2}M/K_{i}$ and $\hat{\alpha}_{i}\triangleq \hat{L}_{i}\log_{2}M/K_{i}$, where $M=2^{q},q=2,3,4,...$
is the order of the QAM modulation. 
\end{defn}
%
\begin{defn}
Assuming that the semantic is correlated among the bits, the semantic information of the $i$th transmitted and received semantic features are defined as, $\alpha_{i}\triangleq L_{i}$, and $\hat{\alpha}_{i}\triangleq \hat{L}_{i}$, respectively. 
\end{defn}}
\hre{
\begin{rem}
For the same semantic feature, it shows different importances for different receivers due to their different interests/tasks/goals. 
\end{rem}}
\end{comment}
{\color{sblue}  This section investigates the semantic-aware power allocation problems for both uncoded forward-with-error and coded discard-with-error schemes, aiming to minimize total power consumption while maintaining  semantic performance. Given that  ultra-low rates are achievable in generative SemCom systems, we focus on highly reliable transmission using uncoded BPSK and finite blocklength coding \cite{polyanskiy2010channel}.} 

\begin{comment}
{\color{red}Considering block fading channels, the channel for the $i$-th semantic data stream  can be expressed by
\begin{equation}
	h_i = \sqrt{P_\mathrm{Loss}}\Tilde{h}_i,
\end{equation} where $P_\mathrm{Loss}=P_{\mathrm{Loss},0}( {d}/{d_0})^{-\alpha}$ is the path loss at distance $d$, with $P_{\mathrm{Loss},0}$ being the path loss at reference distance $d_0$. The term $\Tilde{h}_i$ is Rayleigh fading channel with a variance of $1$. }
\end{comment}

Let $\mathbf{z}_{i}$ be the transmitted signal of the $i$-th  semantic data stream such that $\mathbb{E}\left\{ \mathbf{z}_{i}^{\mathrm{H}}\mathbf{z}_{i}\right\} = Z_i$, where $Z_i$ is the length of the transmitted signal. The $i$-th received semantic signal can be written as: 
\begin{equation}
\mathbf{y}_{i}=h_i\sqrt{q_{i}}\mathbf{z}_{i}+\mathbf{n}_{i},
\end{equation} where $h_{i}$ is the block-fading channel. $\mathbf{n}_{i}$ is the Gaussian noise vector, with each element following the distribution of $ \mathcal{CN}\left(0,\sigma_{i}^{2}\right)$. $q_{i}$ is the allocated power for  the $i$-th semantic data stream. The  signal-to-noise-ratio (SNR) is given by:
\begin{equation}
\mathrm{snr}_{i}=\frac{q_{i}\vert h_i\vert^2}{\sigma_i^2}.
\end{equation} %Due to the fading and noisy characteristic of the wireless channels,  errors occur in the received semantic data streams, and further degrades  the  semantic performance. The error under the uncoded and coded schemes are provided as follows.

%\subsection{Uncoded Forward-with-Error Scheme}
{\color{sblue}Under the channel-uncoded case with BPSK modulation,  the length of the transmit signal satisfies $Z_i=K_i$. The total power consumption is  given by $\sum_{i=1}^I K_iq_{i}$.  The BER is uniform across all bits  under the block-fading channel, i.e., $\psi_{ij}=\psi_i, \forall j=1,\dots, K_i$, which is expressed as:} %(eq. 16)
\begin{equation} 
{\color{sblue}\psi_{i}} = Q(\sqrt{2\mathrm{snr}_i}),\label{eq:BER_p}
%\psi'_{i} = \frac{a_i}{\log_2M_i}Q\left(\sqrt{b_i\mathrm{SNR}_i}\right),\label{eq:BER_p}
\end{equation}where $Q\left(x\right)=\frac{1}{\sqrt{2\pi}}\int_x^{\infty}e^{(-\frac{u^2}{2})}du$ is the Q-function. %Parameters $a_i$ and $b_i$ depend on the adopted the modulation type, which arelisted in \cite[Table 1]{simon2001digital}.  
The  probability of $\hat{\mathbf{K}}_i$ conditioning on $\mathbf{K}_i$ is yielded  by $\mathbb P(\hat{\mathbf{K}}_i|\mathbf{K}_i)=(\psi_i)^{k_i}(1- \psi_i)^{K_i-k_i}$,
where $k_i$ is the number of incorrect bits. 
To minimize total power consumption while ensuring the semantic performance, the problem is formulated as 
\begin{subequations}
\begin{align}
(\mathcal P1): \quad \min_{q_{i}}\quad & \sum_{i=1}^I K_iq_{i}  \label{eq:power_obj}\\
\mathrm{s.t.\quad} & %P^\mathrm{unc}_{\mathrm {forward}}\left(\left\{\Phi_{i}\right\}_i\right)\le\bar{P}\label{eq:power_allocation_ber_f_cons1}.
P_\mathrm{forward}\left(\left\{\color{sblue}{\psi_{i}}\right\}_i\right)\le\bar{P}\label{eq:power_allocation_ber_f_cons1}\\
& \psi_{i} = Q(\sqrt{2\mathrm{snr}_i}),\,\,\forall i\in\mathcal I
%\\& P^{\mathrm{PSNR}}\left(\Psi_{\mathbf{K}_1},\Psi_{\mathbf{K}_2}\right)\le\bar{P}^{\mathrm{PSNR}}, \\
%&P^{\mathrm{MSSSIM}}\left(\Psi_{\mathbf{K}_1},\Psi_{\mathbf{K}_2}\right)\le\bar{P}^{\mathrm{MSSSIM}},\,\, \forall i\in\mathcal I,
\end{align}\label{eq:problem1}\end{subequations}{\color{sblue}where  $\bar{P}$ represents the maximum tolerate perceptual distance.}

%\subsection{Coded Discard-with-Error Scheme}
{\color{sblue}Under the channel-coded case employing finite blocklength coding,  the length of transmit signal  $Z_i$ equals the  channel codeword length $N_{i}$, where $N_{i}\ge K_i$.  The coding rate is given by $K_i/N_i\le 1$, and the total power equals $\sum_{i=1}^I N_iq_{i}$.} According to  \cite{polyanskiy2010channel},  the BLER of the $i$-th semantic data stream is lower bounded by: 
\begin{equation}\label{eq:BLER_polyanskiy}
\Psi_{i}=Q\left(\ln2\sqrt{\frac{N_{i}}{V_{i}}}\left(C_{i}-\frac{K_{i}}{N_{i}}\right)\right),
\end{equation} where $C_{i}$ is the channel capacity expressed as $
    C_{i}=\log_{2}\left(1+\mathrm{snr}_{i}\right)$, $V_{i}$ is the channel dispersion expressed as
$V_i = 1-\left(1+\mathrm{snr}_{i}\right)^{-2}$.
Using this BLER bound, the problem that minimizes total power consumption while ensuring the semantic performance, is formulated as 
\begin{subequations}\label{eq:power_allocation_bler}
\begin{align}
(\mathcal P2): \quad \min_{q_{i}}\quad & \sum_{i=1}^I N_iq_{i}  \label{eq:power_obj}\\
\mathrm{s.t.\quad} & P_\mathrm{discard}\left(\left\{\Psi_{i}\right\}_i\right)\le\bar{P}\label{eq:power_allocation_bler_cons1}\\
& \Psi_{i}=Q\left(\ln2\sqrt{\frac{N_{i}}{V_{i}}}\left(C_{i}-\frac{K_{i}}{N_{i}}\right)\right), \,\, \forall i\in\mathcal I.
\end{align}\label{eq:problem3}\end{subequations}

{\color{sblue}While problems $(\mathcal P1)$ and $(\mathcal P2)$ are challenging to solve optimally due to their non-convex constraints, the following corollary is established based on Corollary \ref{thm:decreasing}.}

\begin{col}\label{lem4}
    The optimal solutions to problems $(\mathcal P1)$ and $(\mathcal P2)$   satisfy the equality conditions in constraints \rm{ (\ref{eq:power_allocation_ber_f_cons1})} and \rm{(\ref{eq:power_allocation_bler_cons1})}, respectively. 
    \begin{proof}
    The perception value is non-increasing with $q_{i}$, as both BER $\psi_{i}$ and BLER $\Psi_i$ are monotonically decreasing with $q_{i}$. Therefore, the optimal solution satisfies the perception constraints with equality.  
    \end{proof}
\end{col}

\section{Semantic-Aware Power Allocation}
%Due the non-convexity of the above investigated problems, it is difficult to obtain the optimal solution. 
Since semantic features exhibit  varying levels of  importance, conventional resource allocation strategies may be inefficient. This section presents two semantic-aware power allocation methods:  a proportional method that decouples the perception constraint to yield a closed-form solution, and a bisection method  that employs bisection search to find a locally optimal solution.

%This section presents two sub-optimal semantic-aware methods  and bisection searching based on Corollary \ref{lem4}. 

%\hre{Start by talking about the challenges of solving P1-3, then say how you are addressing these challenges. Say how many different methods are you proposing. Before you say sub-optimum method, you should say why finding the optimum solution was hard.}

%In this section, we first proposed the semantic-aware proportional method to solve the investigated problems by assuming the independent impacts of the semantic bit streams. In particular to the semantic encoder extracting two semantic features, we proposed a semantic-aware sub-optimal method. 

\subsection{\color{sblue}{Semantic-Aware Proportional Method}}

By assuming the independent impact of  semantic data streams on the perceptual quality of the regenerated signal, we decompose the perception constraint   into $I$ independent constraints on semantic values of the received data streams. Problems $(\mathcal P1)$ and $(\mathcal P2)$
can be  relaxed  into 
\begin{subequations}
\begin{align}
(\mathcal P1\text{-}1):\quad \min_{q_{i}}\quad & \sum_{i=1}^I K_iq_{i} \\
\mathrm{s.t.\quad} &\hat{L}_{i,\mathrm{forward}}\left(\psi_{i}\right) \ge \bar{L}_i,\,\, \forall i\in\mathcal I,\label{eq:P1-1Cons1}
\\& \psi_{i} = Q(\sqrt{2\mathrm{snr}_i}),
\end{align}
\end{subequations}
and
\begin{subequations}
\begin{align}
(\mathcal P2\text{-}1): \, \min_{q_{i}}\quad & \sum_{i=1}^I N_iq_{i} \\ 
\mathrm{s.t.\quad} &\hat{L}_{i,\mathrm{discard}}\left(\Psi_i\right) \ge \bar{L}_i,\,\, \forall i\in\mathcal I, \label{eq:P2-1Cons1}\\
& \Psi_{i}=Q\left(\ln2\sqrt{\frac{N_{i}}{V_{i}}}\left(C_{i}-\frac{K_{i}}{N_{i}}\right)\right), \,\, \forall i\in\mathcal I,
\end{align}
\end{subequations}respectively, where $\bar{L}_i$ denotes the semantic value requirements corresponding to the semantic performance requirement $\bar{P}$. As stated in Remark \ref{rem1},  the semantic value of the received semantic data stream is non-increasing with respect to (w.r.t.) BER $\psi_{i}$ or BLER $\Psi_{i}$. {\color{sblue}Consequently, the optimal solutions to $(\mathcal P1\text{-}1)$ and $(\mathcal P2\text{-}1)$ are achieved when  constraints (\ref{eq:P1-1Cons1}) and (\ref{eq:P2-1Cons1}) are satisfied with equality, establishing the following theorem.}
\begin{thm}\label{lem5}
    {\color{sblue}For problem $(\mathcal P1\text{-}1)$,}  the optimal solutions $q_i^*$ is given by:
    \begin{equation}
     \quad q_{i}^* = \frac{\sigma_{i}^{2}}{2\vert h_{i}\vert^2}\left(Q^{-1}\left(\psi_{i}^*\right)\right)^2,
    \end{equation}
    where $\psi_{i}^*$ is obtained by solving equation $\hat{L}_{i,\mathrm{forward}}\left(\psi_{i}\right) = \bar{L}_i$.  
    
    {\color{sblue}For problem $(\mathcal P2\text{-}1)$,} denote $\Psi_{i}^*$ as the solution to $\hat{L}_{i,\mathrm{discard}}\left(\Psi_{i}\right) = \bar{L}_i$, and define $\alpha_i\triangleq Q^{-1}(\Psi^*_{i})/\sqrt{N_i}$. The optimal solution $q_i^*$ is given by:
    \begin{equation}
    	\quad q_{i}^* = \frac{\sigma_{i}^{2}}{\vert h_{i}\vert^2}\left(e^{\frac{K_i}{N_i}+\eta_i^*}-1\right),
    \end{equation}
    where $\eta_i^*=W(^{2\alpha_i}, ^{-2\alpha_i};-4e^{-2K_i/N_i}\alpha_i^2)/2$, with $W(\cdot)$ denoting the generalized Lambert W function\footnote{The generalized Lambert W function $W(^{t_1}, ^{t_2};a)$ is the solution to the transcendental equation $(x-t_1)(x-t_2)e^x=a$ \cite{mezHo2017generalization}.}. The proof appears in Appendix \ref{Proof_Lemma5}.
\end{thm}

%\hre{Right after proving any theorem/lemma, we should bring some remarks to give some insights from the theorem/lemma. For example, it is not easily clear how e.q. (35) is semantic-aware? If it included $L_i$ in it, then we could give insights on effects of semantic-awareness, etc.}

%It should be noted that the optimal solutions $q_i^*$ to problems $\mathcal P1\text{-}1$ and $\mathcal P2\text{-}1$ are not optimal to original problems $\mathcal P1$ and $\mathcal P2$. 

\subsection{{\color{sblue}Semantic-Aware Bisection Method}}
{\color{sblue}For image tasks, two semantic extractors suffice to obtain semantic features for high-quality image regeneration \cite{lei2023text+}, where one extractor provides  textual description  while the other captures structural features. Given this dual-extractor configuration,  problems $(\mathcal P1)$ and $(\mathcal P2)$ can be reformulated based on Corollary \ref{lem4} as follows. }

For the uncoded BPSK scheme,  the allocated power derived from  \eqref{eq:BER_p} can be expressed as $p_i=\frac{ \sigma_{i}^{2}}{2\vert h_{i}\vert^2}(Q^{-1}({\psi_i}))^2$.  Thus, problem $(\mathcal P1)$ with dual-extractor configuration is reduced into: 
\begin{subequations}
	\begin{align}
		(\mathcal P1\text{-} 2): \quad \quad\min_{{\psi_{1}}, {\psi_{2}}}\quad & \sum_{i=1}^{2}\frac{ K_i\sigma_{i}^{2}}{2\vert h_{i}\vert^2}\left(Q^{-1}\left({\psi_i}\right)\right)^2 \\
		\mathrm{s.t.\quad} &  P_{\mathrm {forward}}(\psi_{1}, \psi_{2})=\bar{P}.\label{eq:pro1-2_cons1}
	\end{align}\label{eq:problem0}\end{subequations}
For the finite blocklength coding scheme, the allocated power  derived from \eqref{eq:BLER_polyanskiy} is given by $p_i=\frac{\sigma_i^2}{\vert h_{i}\vert^2} \mathrm{snr}_i(\Psi_i)$, where $\mathrm{snr}_i(\Psi_i)$ is the solution to  equation \eqref{eq:BLER_polyanskiy}. Accordingly, problem $(\mathcal P2)$ with dual-extractor configuration is reduced into:
 
\begin{subequations}
\begin{align}
(\mathcal P2\text{-} 2): \quad \min_{\Psi_{1}, \Psi_{2}}\quad & \sum_{i=1}^2\frac{N_i\sigma_i^2}{\vert h_{i}\vert^2} \mathrm{snr}_i(\Psi_i) \\
\mathrm{s.t.\quad} &  P_{\mathrm {discard}}\left(\Psi_{1}, \Psi_{2}\right)=\bar{P}.\label{eq:pro3-2_cons1}
\end{align}\label{eq:problem0}\end{subequations}

\begin{comment}
Denote the objectives of problems $\mathcal P1\text{-}2$, $\mathcal P2\text{-}2$ and $\mathcal P3\text{-}2$ as $f_1$, $f_2$ and $f_3$, respectively. Their gradients w.r.t. the optimized variables are given by
\begin{subequations}

\begin{align}
    &\nabla_{\psi_{i}}f_1 
   =-\frac{K_i\sigma_i^2}{\sqrt{2\pi}\vert h_i\vert^2}Q^{-1}\left(\psi_{i}\right)e^{-\frac{\psi_{i}^2}{2}},\\
   &\nabla_{\Psi_{i}}f_2=-\frac{\sigma_i^2\left(1-\Psi_{i}\right)^{\frac{1}{K_i}-1}}{\sqrt{2\pi}\vert h_i\vert^2}Q^{-1}\left(\psi'_{i}\right)e^{-\frac{\left(\psi'_{i}\right)^2}{2}},\\
   &\nabla_{\Psi_{i}}f_3=\frac{K_1\sigma_1^2}{\vert h_{1}\vert^2}\frac{\partial \mathrm{SNR}_i}{\partial \Psi_{i}}=\frac{K_i\sigma_i^2}{\vert h_{i}\vert^2}\frac{g_\mathrm{num}\left(\mathrm{SNR}_{i}\right)}{g_\mathrm{den}\left( \Psi_i\right)},
\end{align}\end{subequations}where $\psi'_{i}=1-\left(1-\Psi_{i}\right)^{\frac{1}{K_i}}$.  $g_\mathrm{den}\left(\gamma_i\right)$ and $g_\mathrm{num}$ are the gradients given by  
\begin{subequations}
\begin{align}
&g_\mathrm{den}\left(\gamma_i\right)=\left(\ln2\sqrt{\frac{n_{i}}{1-\left(1+\gamma_{i}\right)^{-2}}}\left(\log_{2}\left(1+\gamma_{i}\right)-\frac{K_{i}}{n_{i}}\right)\right)',
\\
&g_\mathrm{num} = \left(Q^{-1}\left(\Psi_{i}\right)\right)'=-\frac{1}{\sqrt{2\pi}}e^{-\frac{\left(Q^{-1}\left(\Psi_{i}\right)\right)^{2}}{2}}.
\end{align}
\end{subequations}

\hre{The gradient of $\nabla_{\psi_1}\psi_2$ and $\nabla_{\Psi_1}\Psi_2$  has}
\end{comment}

The local optimal solution can be obtained through bisection search by locating where the objective function gradient equals zero. For simplicity of notation, denote $\left(\Phi_{1},\Phi_{2}\right)$ ($\Phi_{1}\in\{{\psi_{1}}, {\Psi_{1}}\}$, $\Phi_{2}\in\{{\psi_{2}}, {\Psi_{2}}\}$) as the optimizing variables,  and   $f$ ($f\in\{f_1, f_2\}$) as the objective functions  of the above problems. The feasible solutions $(\Phi_{1},\Phi_{2})$ forms a line on the perception-error surfaces. For any two feasible solutions $(\Phi^{(1)}_{1}, \Phi^{(1)}_{2})$ and $(\Phi^{(2)}_{1}, \Phi^{(2)}_{2})$,  we have $\Phi^{(2)}_{2}\le\Phi^{(2)}_{2}$ if $\Phi^{(1)}_{1}\ge\Phi^{(2)}_{1}$.  
Denoting the line endpoints as $\left(\Phi_{1}^L,\Phi_{2}^L\right)$ and $\left(\Phi_{1}^R,\Phi_{2}^R\right)$, where $\Phi_{1}^R\ge\Phi_{1}^L$, 
 the procedure to obtain the local optimal solution is summarized in  Algorithm \ref{alg1}. 
\begin{algorithm}[t]
	%\textsl{}\setstretch{1.8}
	\renewcommand{\algorithmicrequire}{\textbf{Input:}}
	\renewcommand{\algorithmicensure}{\textbf{Output:}}
	\caption{{\color{sblue}Semantic-Aware Bisection  Method for  Semantic  Encoder with Dual-Extractor Configuration}} 
	\label{alg1}
	 Input:   $(\Phi_{1}^L,\Phi_{2}^L)$,  $(\Phi_{1}^R,\Phi_{2}^R)$.\\
	Output:  $(\Phi_{1},\Phi_{2})$.
	
	\begin{algorithmic}[1]

        \STATE Compute partial gradients     $(\frac{\partial f}{\partial \Phi_{1}^L}, \frac{\partial f}{\partial \Phi_{2}^L})$ and $(\frac{\partial f}{\partial \Phi_{1}^R}, \frac{\partial f}{\partial \Phi_{2}^R})$.
         \STATE Compute gradients  $\nabla_{\Phi_1^L}\Phi_2^L$ and $\nabla_{\Phi_1^R}\Phi_2^R$ by implicit differentiation of (\ref{eq:pro1-2_cons1}) or (\ref{eq:pro3-2_cons1}).
         \STATE \textbf{if} $\frac{\partial f}{\partial \Phi_{1}^L}+\nabla_{\Phi_1^L}\Phi_2^L\frac{\partial f}{\partial \Phi_{2}^L}\ge 0$.
         \STATE \quad $(\Phi_{1},\Phi_{2})\leftarrow (\Phi_{1}^L,\Phi_{2}^L)$.
         
         \STATE \textbf{else if}
         $\frac{\partial f}{\partial \Phi_{1}^R}+\nabla_{\Phi_1^R}\Phi_2^R\frac{\partial f}{\partial \Phi_{2}^R}\le 0$
         \STATE \quad $(\Phi_{1},\Phi_{2})\leftarrow (\Phi_{1}^R,\Phi_{2}^R)$.
         \STATE \textbf{else}
         \STATE \quad \textbf{while} $\Phi_{1}^R-\Phi_{1}^L \ge \epsilon$
		\STATE \quad \quad $\Phi_1=(\Phi_{1}^R+\Phi_{1}^L)/2$.
  
		\STATE \quad  \quad Obtain $\Phi_2$ by solve the equation (\ref{eq:pro1-2_cons1}) or (\ref{eq:pro3-2_cons1}).

            \STATE  \quad \quad Compute partial gradients     $(\frac{\partial f}{\partial \Phi_{1}}, \frac{\partial f}{\partial \Phi_{2}})$ and  $\nabla_{\Phi_1}\Phi_2$.
            
             \STATE  \quad\quad \textbf{if} $\frac{\partial f}{\partial \Phi_{1}}+\nabla_{\Phi_1}\Phi_2\frac{\partial f}{\partial \Phi_{2}}\ge 0$.
             
            \STATE \quad\qquad$(\Phi_{1}^R, \Phi_{2}^R)\leftarrow (\Phi_1, \Phi_2)$.
            
            \STATE  \quad\quad \textbf{else} 
            \STATE  \quad\qquad$(\Phi_{1}^L, \Phi_{2}^L)\leftarrow (\Phi_1, \Phi_2)$.
            \STATE  \quad\quad \textbf{end} 
		\STATE \quad\textbf{end}  
		\STATE  \textbf{end}
            %\STATE  Compute the objective function $f$ 
		%\ENSURE  Total power consumption $f$ %$\left(\Phi_1, \Phi_2\right)$ 
	\end{algorithmic}  
\end{algorithm}

\section{Simulation Results\label{sec:VI}}
This section considers  the image task and focuses on three key aspects: 1). Validating the effectiveness of the proposed generative SemCom framework; 2). Verifying perception-error functions and the semantic values of different semantic data streams; and 3). Assessing the performance of the semantic-aware power allocation methods. For comprehensive evaluation, both CLIP  and  MS-SSIM metric are considered for semantic performance assessment. % to measure the impact of $\mathcal T$ on the distortion performance. %The analysis is the same by replacing the perceptual distance, leading to a distortion-error function.

\begin{figure*}[tp]
\vspace{1em}
\centering
\includegraphics[width=0.9\textwidth]{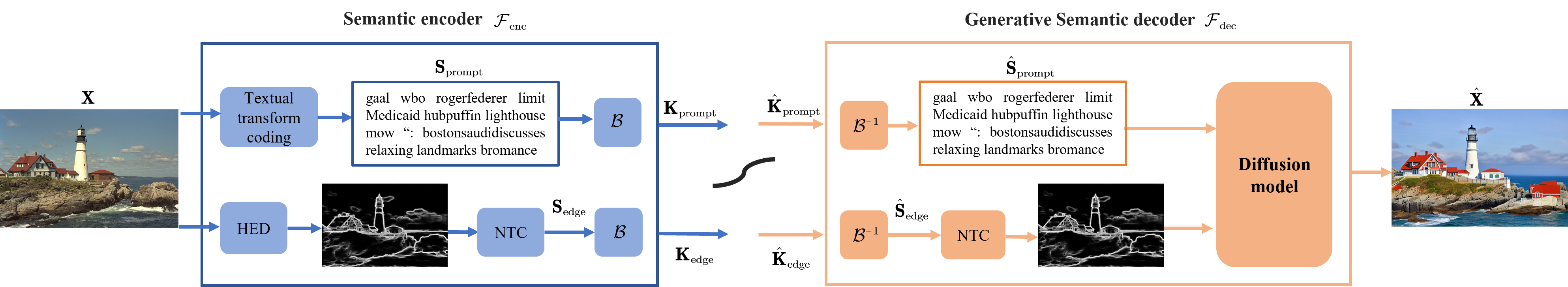}
\caption{{\color{sblue}The proposed generative SemCom framework for the image task with dual semantic extractors for textual prompt and edge map features.} }
\label{fig:system_image}
\end{figure*}

\begin{figure}[t]
	\centering
	\includegraphics[width=0.9\columnwidth]{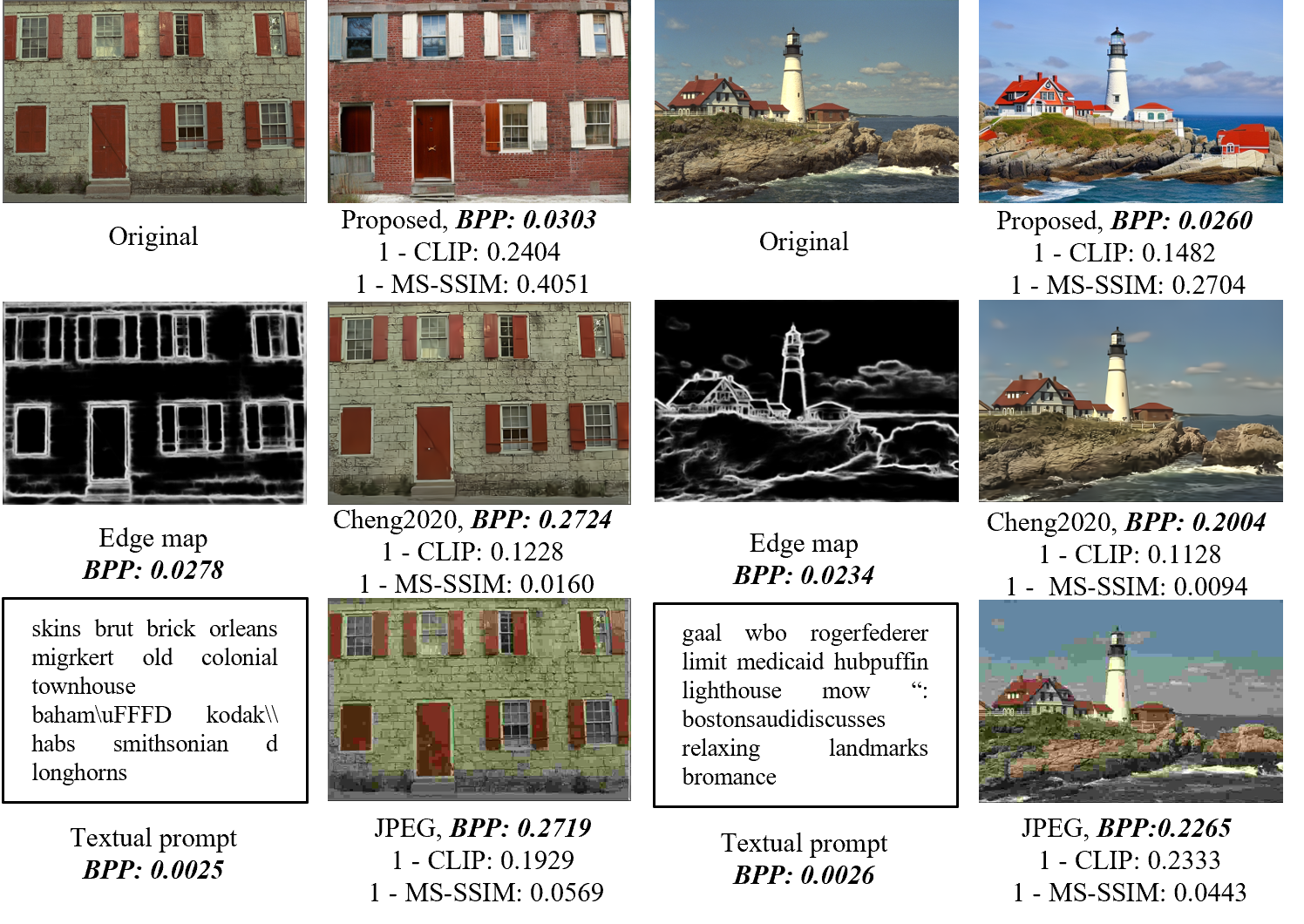}
	\caption{{Visual quality and compression rate comparison of the proposed approach against JPEG and Cheng2020.}}
	\label{fig:compression_rate}
\end{figure}

\begin{figure*}[th]
	\centering
	\subfigure[]{
		\includegraphics[width=0.29\textwidth]{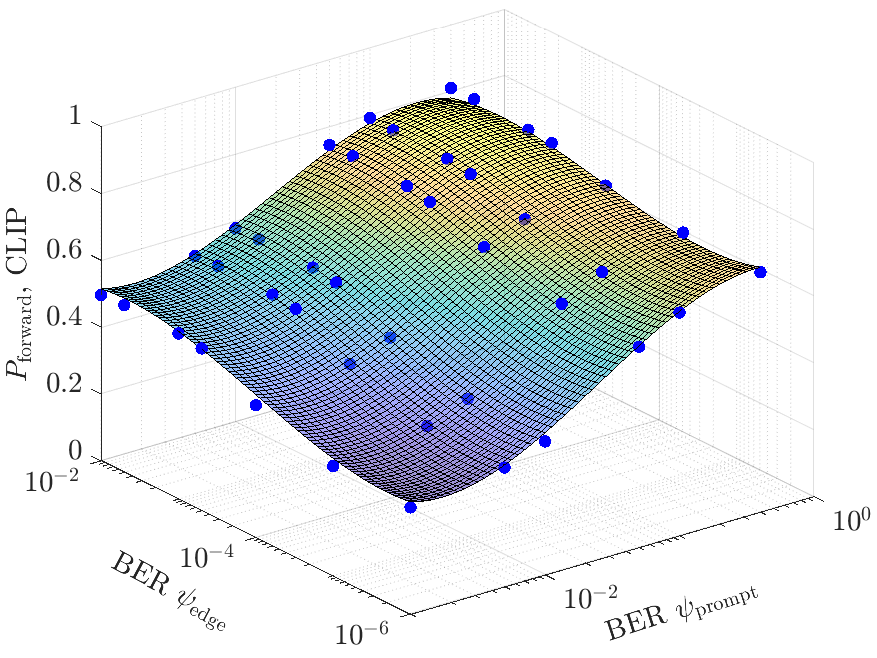}}
	\qquad
	\subfigure[]{
		\includegraphics[width=0.29\textwidth]{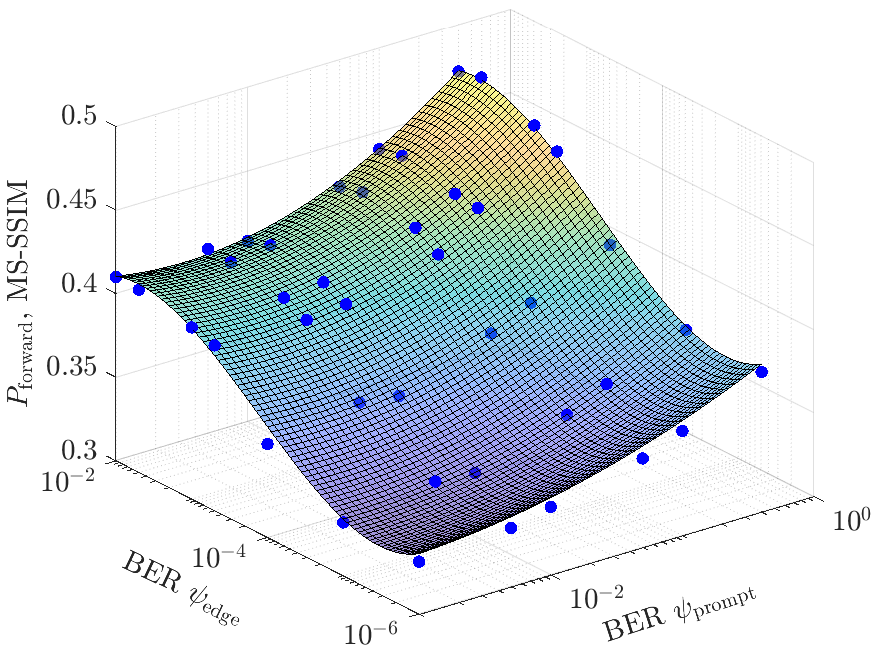}}
	\quad
	\subfigure[]{
		\includegraphics[width=0.29\textwidth]{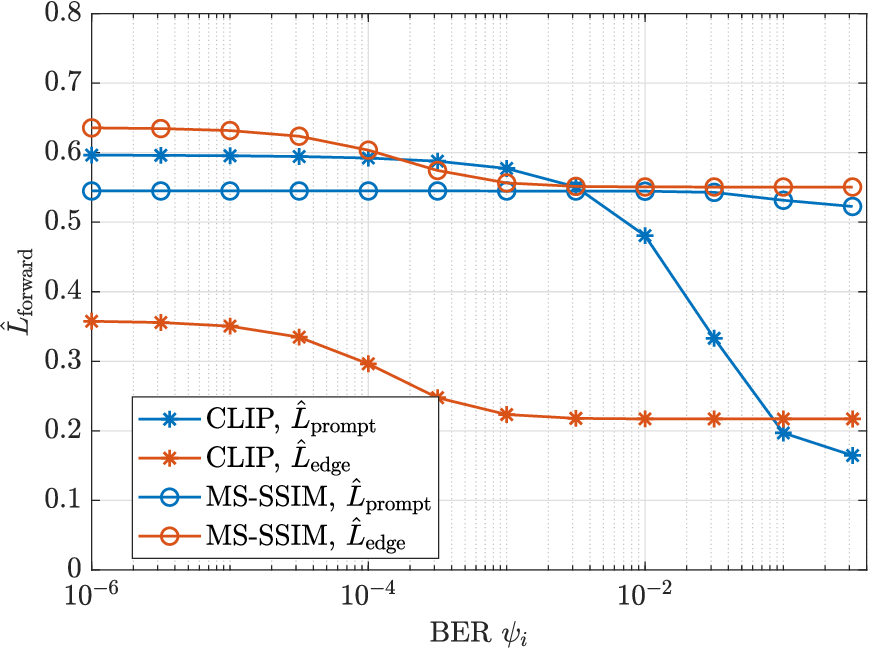}}
	
	\subfigure[]{\includegraphics[width=0.29\textwidth]{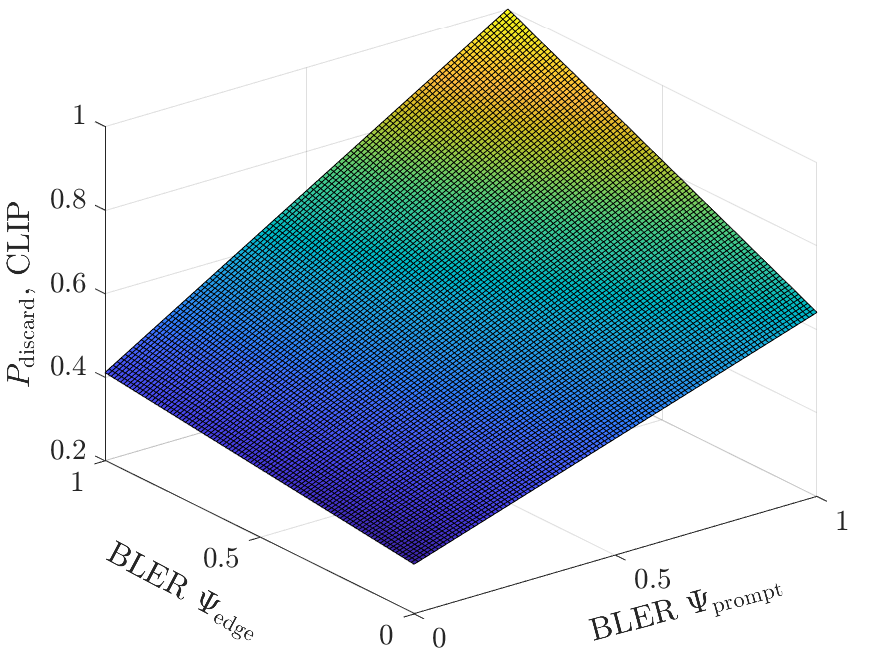}}
	\qquad
	\subfigure[]{\includegraphics[width=0.29\textwidth]{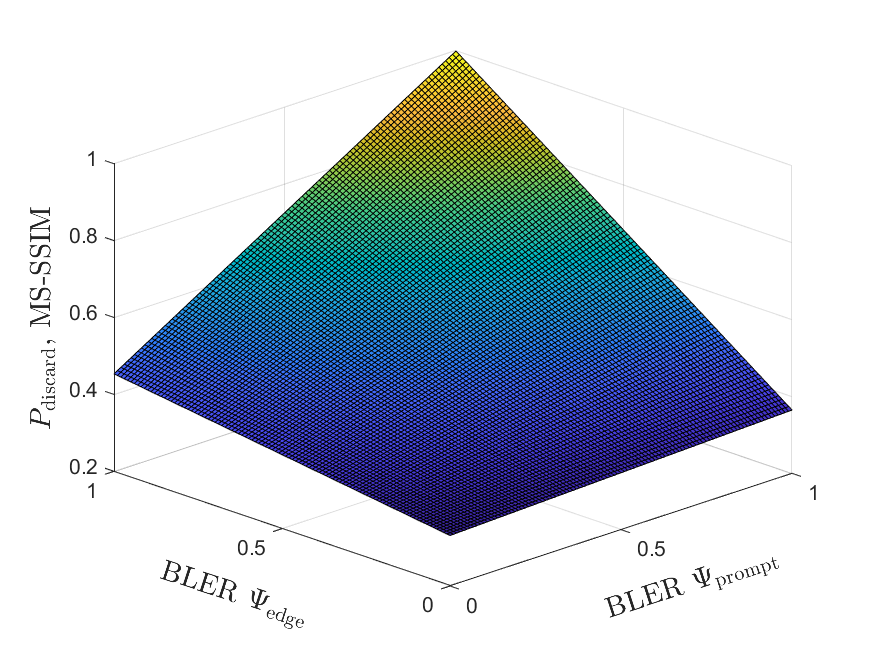}}
	\qquad
	\subfigure[]{\includegraphics[width=0.29\textwidth]{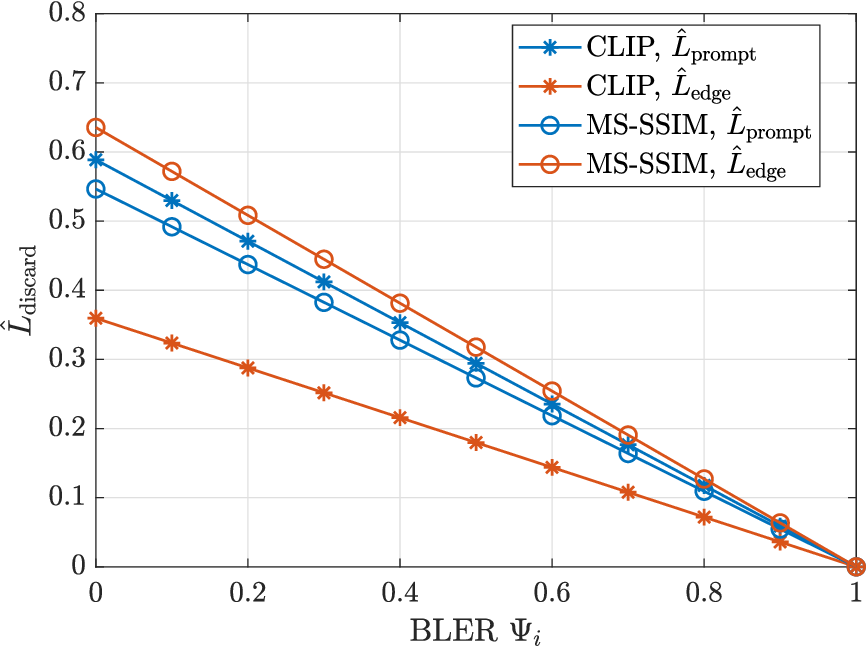}}
	\caption{{\color{sblue}The perception-error functions and semantic values: (a-c). Uncoded forward-with-error scheme. (e-f). Coded discard-with-error scheme.}}	
	\label{fig:P_E_function}
\end{figure*}

\begin{comment}
\begin{figure*}[th]
\centering
\subfigure[]{
\includegraphics[width=0.33\textwidth]{figures/fit_text_sketch_ber.eps}}
%\subfigure[]{\includegraphics[width=0.32\textwidth]{figures/fit_text_sketch_ber_clip_discard.eps}}
\qquad
\subfigure[]{\includegraphics[width=0.33\textwidth]{figures/fit_text_sketch_bler.eps}}
\hspace{2em}
\subfigure[]{
\includegraphics[width=0.33\textwidth]{figures/fit_text_sketch_ber_ssim.eps}}
%\subfigure[]{\includegraphics[width=0.32\textwidth]{figures/fit_text_sketch_ber_ssim_discard.eps}}
\qquad
\subfigure[]{\includegraphics[width=0.33\textwidth]{figures/fit_text_sketch_bler_ssim.eps}}
\caption{{\color{sblue}The perception-error functions: (a). Uncoded forward-with-error scheme with CLIP metric. (b). Coded discard-with-error scheme with CLIP metric. (c). Uncoded forward-with-error scheme with MS-SSIM metric. (d). Coded discard-with-error scheme with MS-SSIM metric.}}
\label{fig:P_E_function}
\end{figure*}
\end{comment}
\begin{comment}
\begin{table}[t]
\caption{Parameter settings for wireless transmission}
\centering
\begin{tabular}{ccc}
\hline
Parameters& values\\
\hline
Distance $d$ & $100$\,m\\
Reference distance $d_0$ & $1$\,m\\
Path loss at the reference distance $h_0$ & $-30$\,dB\\
Path loss exponent $\alpha$ & $-3.4$\\
Noise power $\sigma_i^2$ & $-110$\,dBm\\
Channel coding rate $K_i/N_i$& $0.8$\\
\hline
\end{tabular}
\end{table}
\end{comment}
\subsection{Effectiveness of the Generative SemCom Framework}
Fig. \ref{fig:system_image} illustrates the proposed generative SemCom framework for the image task. {The semantic encoder utilizes two extractors: a textual prompt feature extractor using textual transform coding via prompt inversion  \cite{wen2023hard}, and an edge map feature extractor employing HED with NTC \cite{balle2020nonlinear}  for compression.} For the semantic decoder, the pre-trained ControlNet  \cite{zhang2023adding} built upon the Stable Diffusion model \cite{rombach2022high}  is adopted. The channel for the $i$-th semantic data stream  is modeled as
$ h_i = \sqrt{P_\mathrm{Loss}}\Tilde{h}_i$. Here,  the path loss is given by $P_\mathrm{Loss}=P_{\mathrm{Loss},0}( {d}/{d_0})^{-\alpha}$, with distance $d=100 \, \text{m}$, reference path loss $P_{\mathrm{Loss},0}=-30\,\text{dB}$ at $d_0=1\, \text{m}$, and path loss exponent $\alpha=-3.4$. $\Tilde{h}_i$ is Rayleigh fading channel with a variance of $1$. The noise power is set to $\sigma_i^2=-110\,$dBm. The coding rate is set to $K_i/N_i=0.8$ for the coded discard-with-error scheme. {Due to the complexity of deriving explicit perception-error functions, numerical simulations are conducted using the Kodak dataset \cite{Kodak}.}    For notation simplicity, we  occasionally use subscripts $1$ and $2$ to represent the textual prompt and edge map features, respectively.

%Given the pre-trained AI models based semantic encoder $\mathcal F_{en}$, semantic decoder $\mathcal F_{de}$, and binary scheme $\mathcal B_i$, the semantic performance can be expressed by

%\hre{The text and sketch prompts are considered as two semantic features with different amounts of semantic units relative to the task/interest of the receiver. }  Since the AI models used in the semantic encoder is pre-trained without further fine-tuning, the obtained lossy compression rate can be written as  $K_1+K_{2}$, where $K_1$ and $K_{2}$ are the lengths of text and sketch semantic bit streams, respectively. 

Fig. \ref{fig:compression_rate} illustrates the regenerated images using the proposed generative SemCom framework, compared against traditional JPEG \cite{wallace1992jpeg} and the deep learning-based  Cheng2020 \cite{cheng2020learned} compression approaches. The compression rate is measured by bit per pixel (bpp). The results demonstrate that our approach achieves significantly lower compression rates   ($0.0303$ bpp and $0.0260$ bpp) in the two example images, outperforming both JPEG ($0.2719$ bpp and $0.2265$ bpp) and Cheng2020 ($0.2724$ bpp and $0.2004$ bpp) by approximately a factor of $10$.  This significant compression improvement maintains comparable perceptual quality, particularly when evaluated using the CLIP metric. These visual results validate the effectiveness of the proposed generative SemCom framework in achieving ultra-low rate while preserving semantics. For comprehensive evaluation of transmission reliability effects, Fig. \ref{fig:examples} in Appendix \ref{generated_image} presents additional regenerated images across varying BER $\psi_i$ and BLER $\Psi_i$. It shows that showing that the visual qualities of the regenerated images are degrading with  transmission errors. Moreover, the results under the coded discard-with-error scheme demonstrate that the textual prompt feature better provides the semantic contents while the edge map captures the structural similarity.

\subsection{Perception-Error Function and Semantic Value} 
Fig. \ref{fig:P_E_function} illustrates the perception-error functions and semantic values of data streams for both uncoded forward-with-error and coded discard-with-error schemes. {\color{sblue} For the uncoded forward-with-error scheme, the perception-error functions are derived through curve fitting of numerical simulation data shown as dots. Without transmission errors, both schemes achieve the best-case perception values ($P_\mathrm{best}$), which are $0.3191$ and  $0.3313$ for CLIP, and MS-SSIM metrics, respectively. With maximum transmission errors, the worst-case values ($P_\mathrm{worst}$) reach $0.8112$ for the CLIP and $0.4720$ for the MS-SSIM metrics under the uncoded scheme, and $1$ under the coded scheme. As the increase of  BERs and BLERs, the perception qualities of the regenerated images are degraded, confirming {Corollary \ref{thm:decreasing}}. Notably, prompt features demonstrate stronger influence on CLIP performance, while edge map features show greater impact on MS-SSIM metrics,}  as CLIP measures similarity in high-level meaning while MS-SSIM captures spatial structural similarity.  In addition, the edge map feature is more vulnerable to BER than the textual prompt feature, due to further NTC compression and larger data stream length.

\begin{comment}
\begin{figure*}[tbp]
\centering
\subfigure[]{
\includegraphics[width=0.38\textwidth]{figures/semantic_value_bpsk_ber.eps}}
%\subfigure[]{\includegraphics[width=0.32\textwidth]{figures/semantic_value_bpsk_bler.eps}}
\qquad\quad
\subfigure[]{\includegraphics[width=0.38\textwidth]{figures/semantic_value_bler.eps}}
\caption{{\color{sblue}Semantic values of textual prompt and edge map semantic data streams in terms of CLIP and MS-SSIM metrics. (a). Uncoded forward-with-error scheme. (b). Coded discard-with-error scheme.}}
\label{fig:semantic values}
\end{figure*}
\end{comment}

\begin{figure}[tbp]
	\centering
	\hspace{0.015em}
	\subfigure[]{\includegraphics[width=0.9\columnwidth]{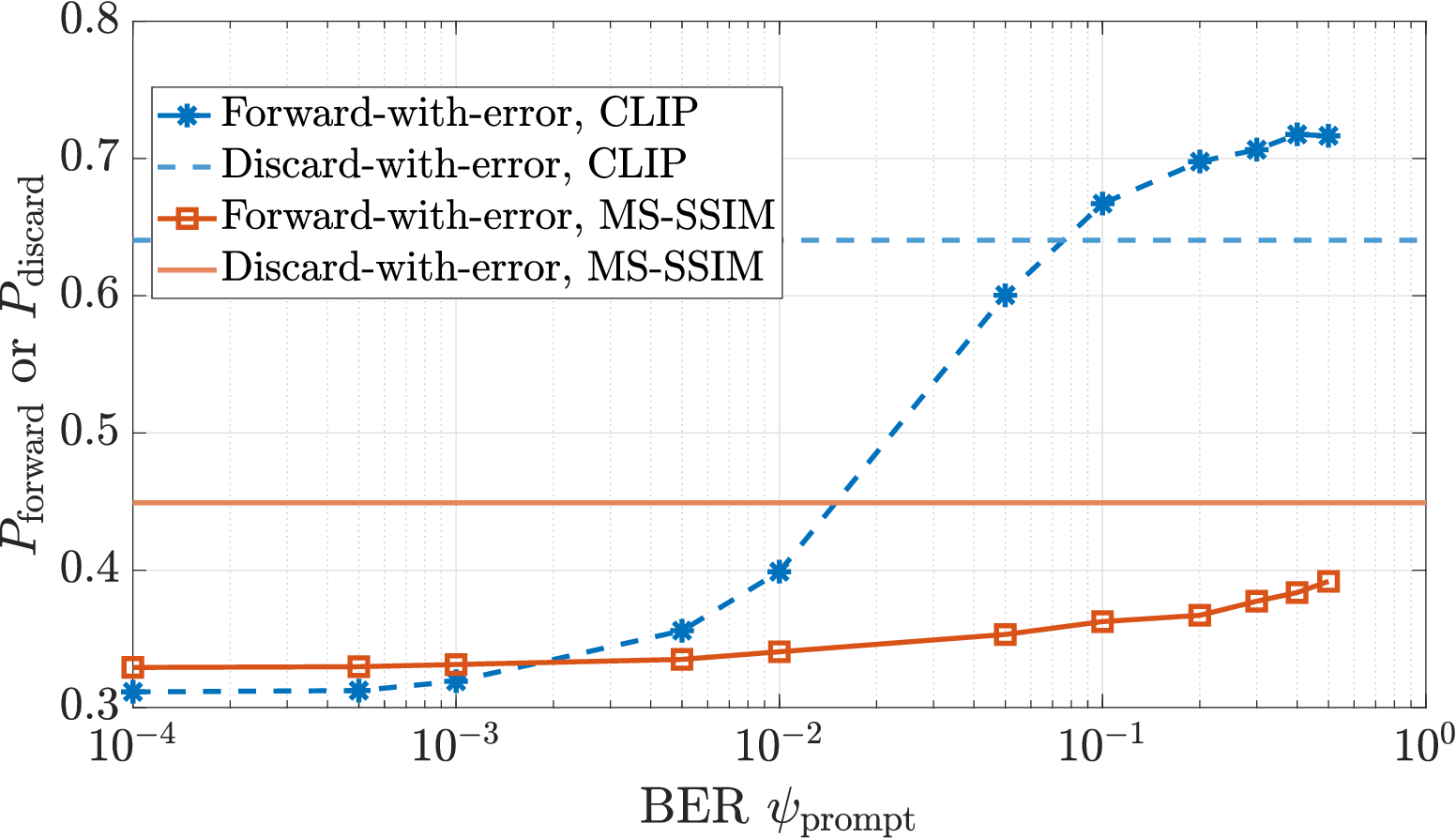}
	}
	\subfigure[]{\includegraphics[width=0.9\columnwidth]{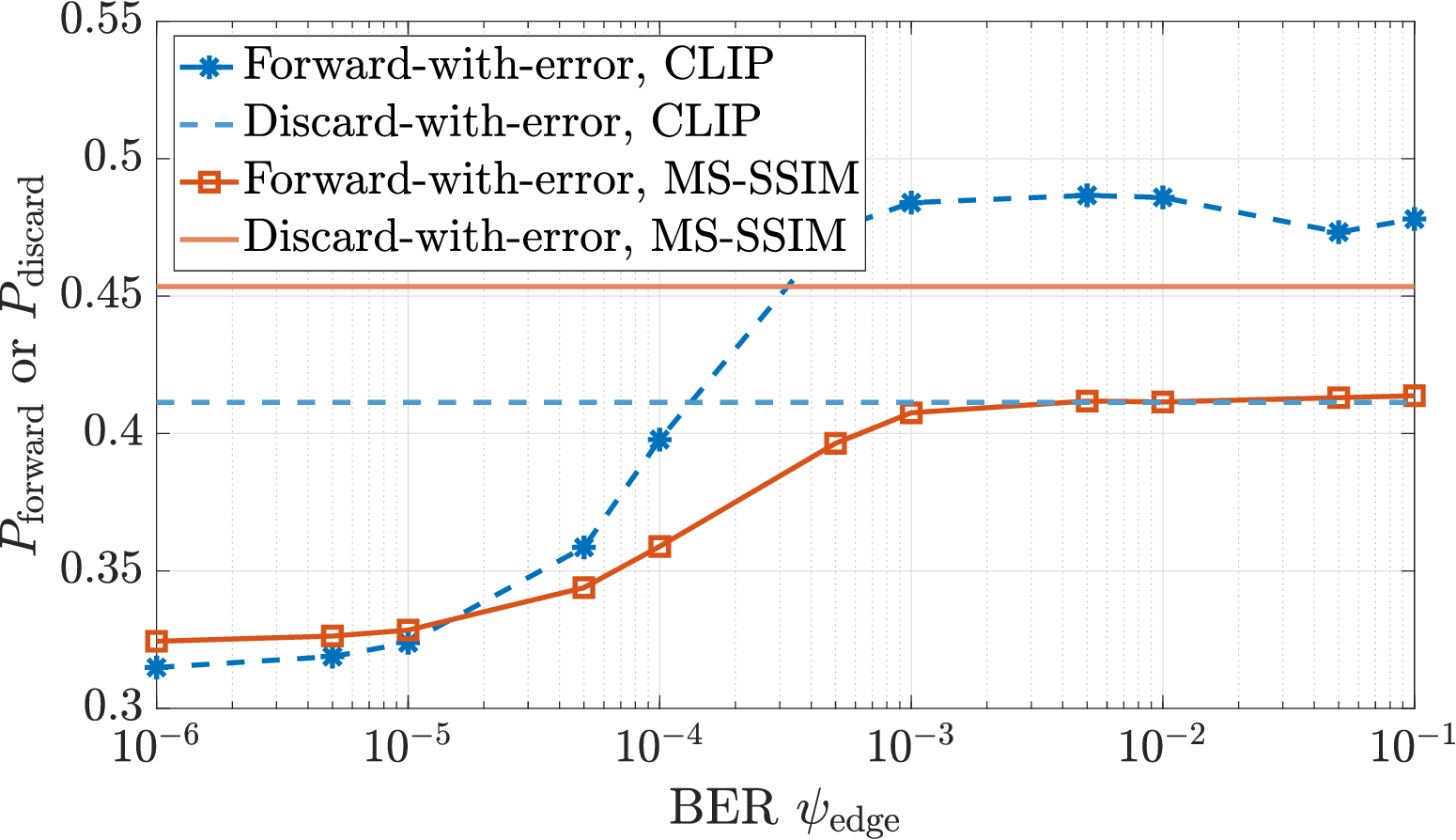}}
	\caption{\color{sblue}The perception-error functions in the case where one of semantic features is correctly received.}
	\label{fig:comparison_discard}
\end{figure}

\begin{comment}
\begin{figure}[tbp]
\centering
\hspace{0.015em}
\subfigure[]{\includegraphics[width=0.455\columnwidth]{figures/comparison_clip_discard_edge.eps}
}
\subfigure[]{\includegraphics[width=0.47\columnwidth]{figures/comparison_clip_discard_prompt.eps}}
\subfigure[]{
\includegraphics[width=0.47\columnwidth]{figures/comparison_msssim_discard_edge.eps}}
\subfigure[]{
\includegraphics[width=0.47\columnwidth]{figures/comparison_msssim_discard_prompt.eps}}
\caption{The perception-error function with/without prompt/edge map}
\label{fig:comparison_discard}
\end{figure}
\end{comment}

The semantic values are also decreasing with BERs or BLERs. For the CLIP metric, textual prompt and edge map data streams show semantic values of $L_1=0.5887$ and $L_2=0.3596$ respectively, while for the MS-SSIM metric, they show values of $L_1=0.5465$ and $L_2=0.6355$. The relationship $L_1 + L_2 > 1-P_\mathrm{best}$ indicates that prompt and edge map semantic features are not semantically independent.  The behavior of semantic values differs between the two error-handling schemes.  Fig. \ref{fig:P_E_function}(c) shows that under uncoded forward-with-error scheme, the received semantic data streams maintain positive semantic values even at maximum BER. However, the semantic values under coded discard-with-error scheme approach $0$ as BLER nears $1$ as shown in Fig. \ref{fig:P_E_function}(f), as the semantic decoder becomes inactive when all received streams contain errors and are discarded. {\color{sblue}  Fig. \ref{fig:comparison_discard} shows the perception-error functions in the case of one	
correctly received	semantic feature,  revealing that forwarding additional semantic data stream with high BER degrades semantic performance compared to discarding them at the decoder. This finding validates the proposed discard-with-error scheme.
}

\begin{figure*}[tbp]
\centering
\subfigure[]{
\includegraphics[width=0.45\textwidth]{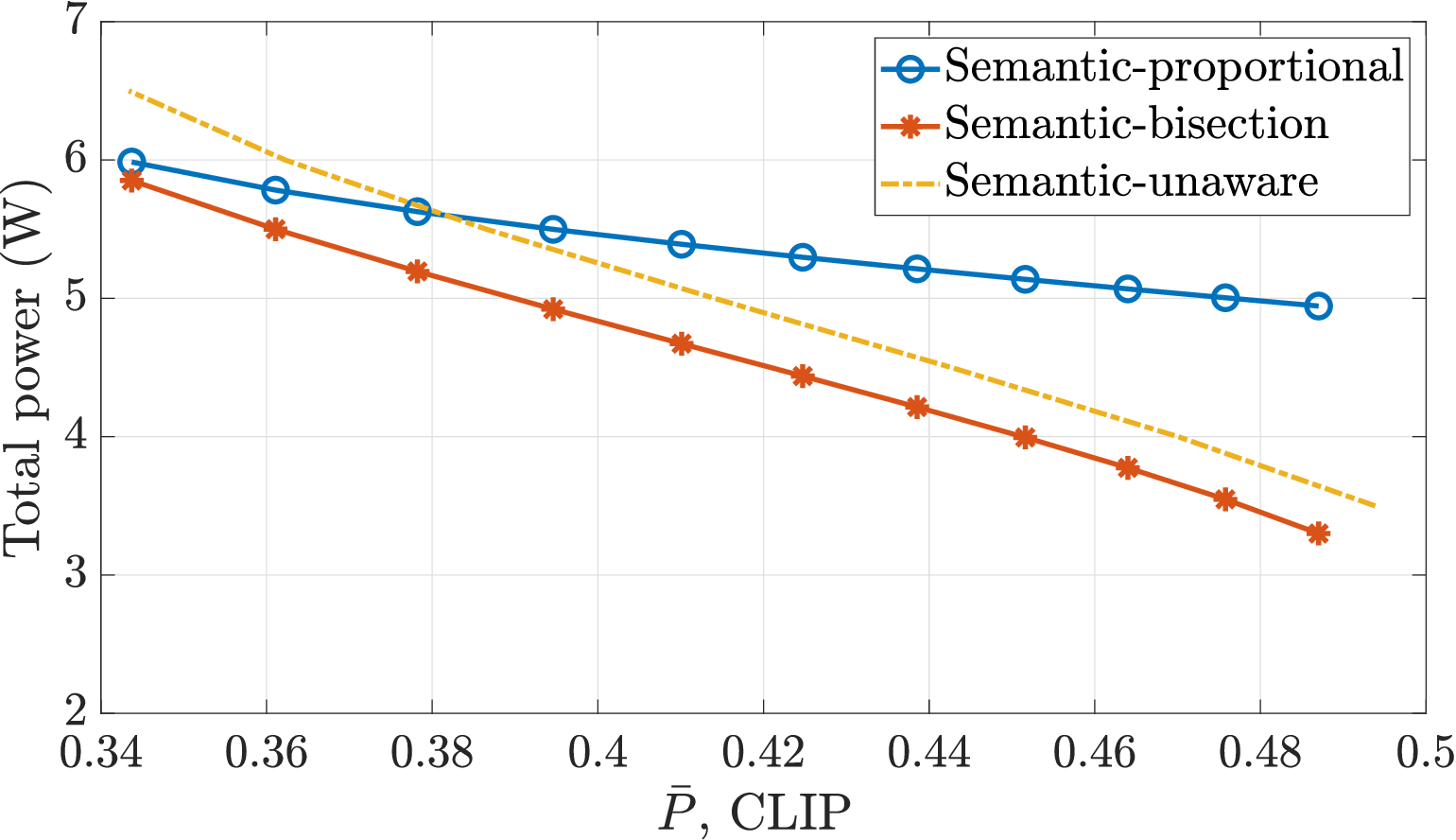}}
%\subfigure[]{\includegraphics[width=0.32\textwidth]{figures/power_ber_bpsk_discard.eps}}
\qquad
\subfigure[]{\includegraphics[width=0.45\textwidth]{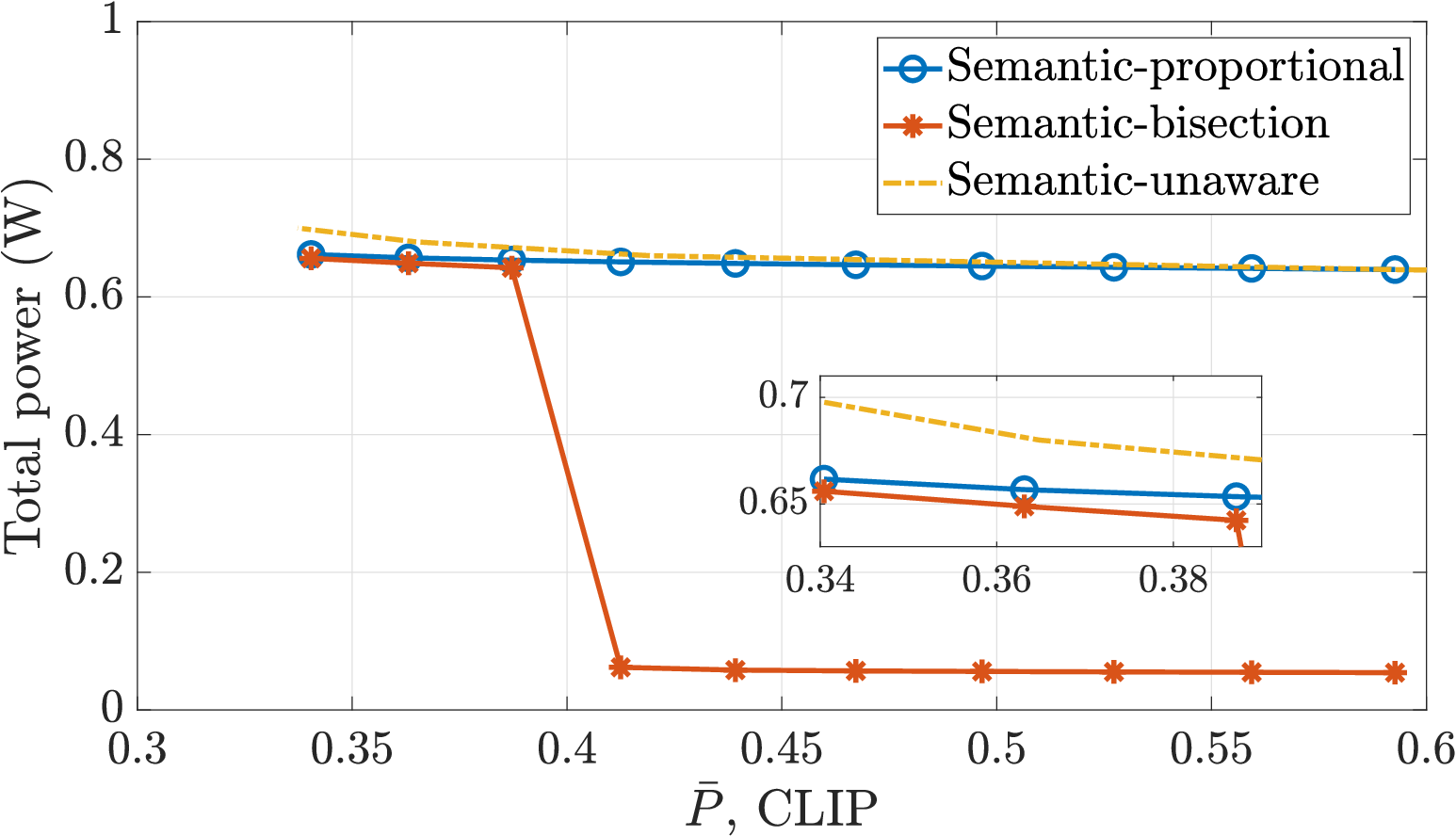}}
\hspace{2em}
\subfigure[]{
\includegraphics[width=0.45\textwidth]{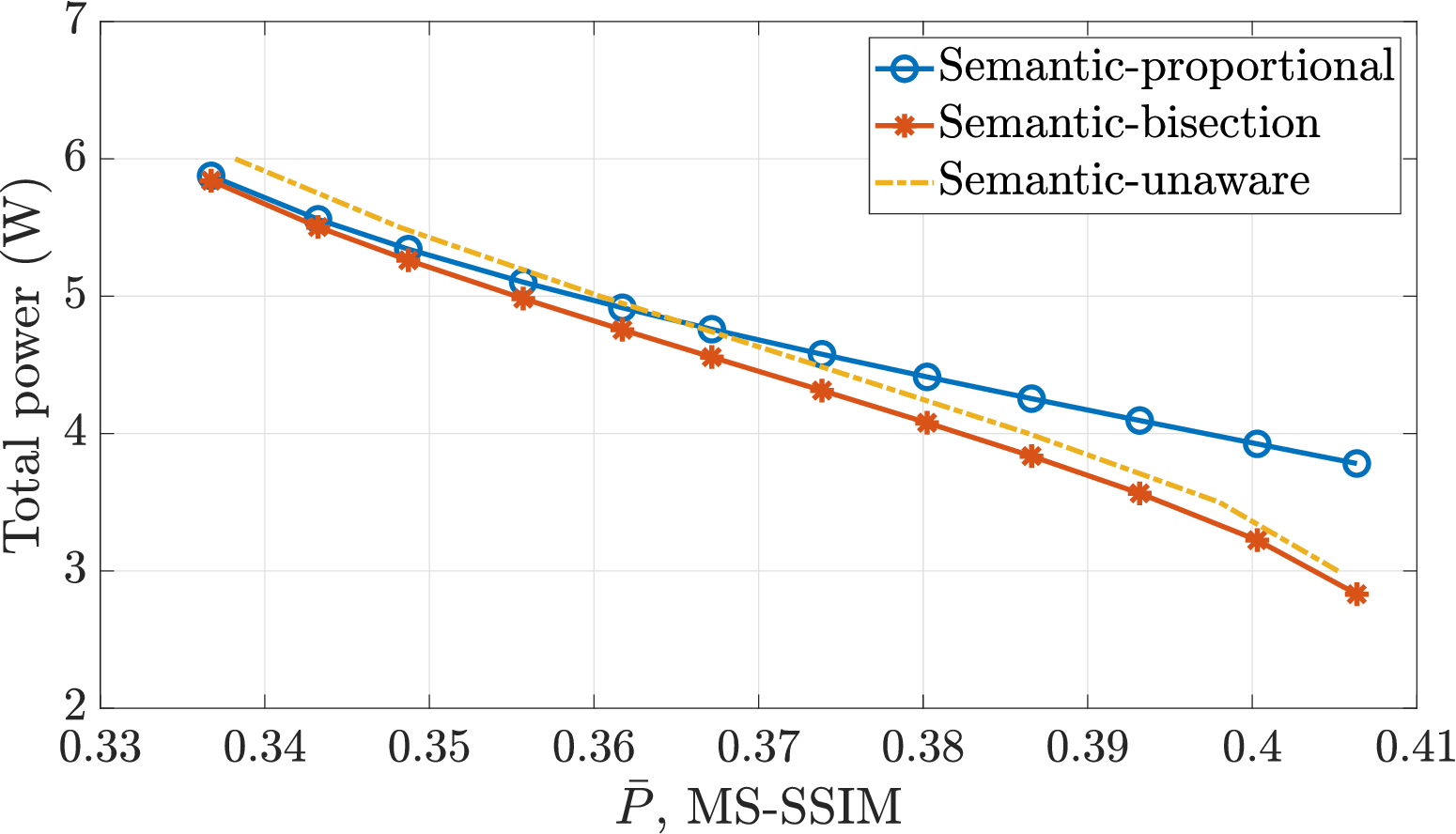}}
%\subfigure[]{\includegraphics[width=0.45\textwidth]{figures/power_ber_bpsk_discard_msssim.eps}}
\qquad
\subfigure[]{\includegraphics[width=0.45\textwidth]{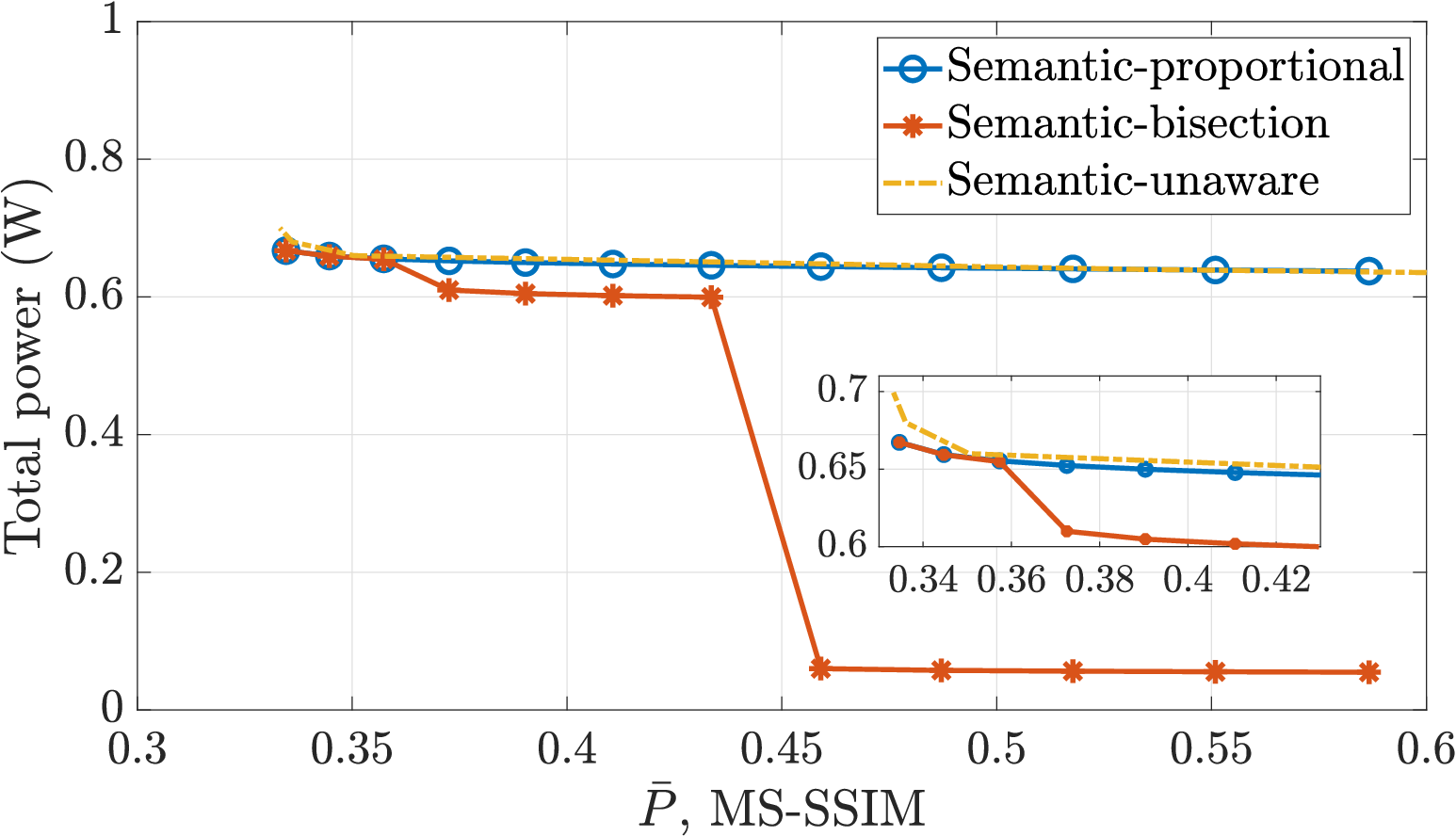}}
\caption{{\color{sblue}Comparisons of total power consumption. (a). Uncoded forward-with-error with the CLIP metric. (b) Coded discard-with-error with the CLIP metric. (c). Uncoded forward-with-error with the MS-SSIM metric. (d) Coded discard-with-error with the MS-SSIM metric.}}
\label{fig:power_barP}
\end{figure*} 
\begin{figure}[tbp]
	\centering
	%\subfigure[]{\includegraphics[width=0.45\columnwidth]{figures/power_bit_uncoded_forward_clip.eps}}
	\subfigure[]{
		\includegraphics[width=0.9\columnwidth]{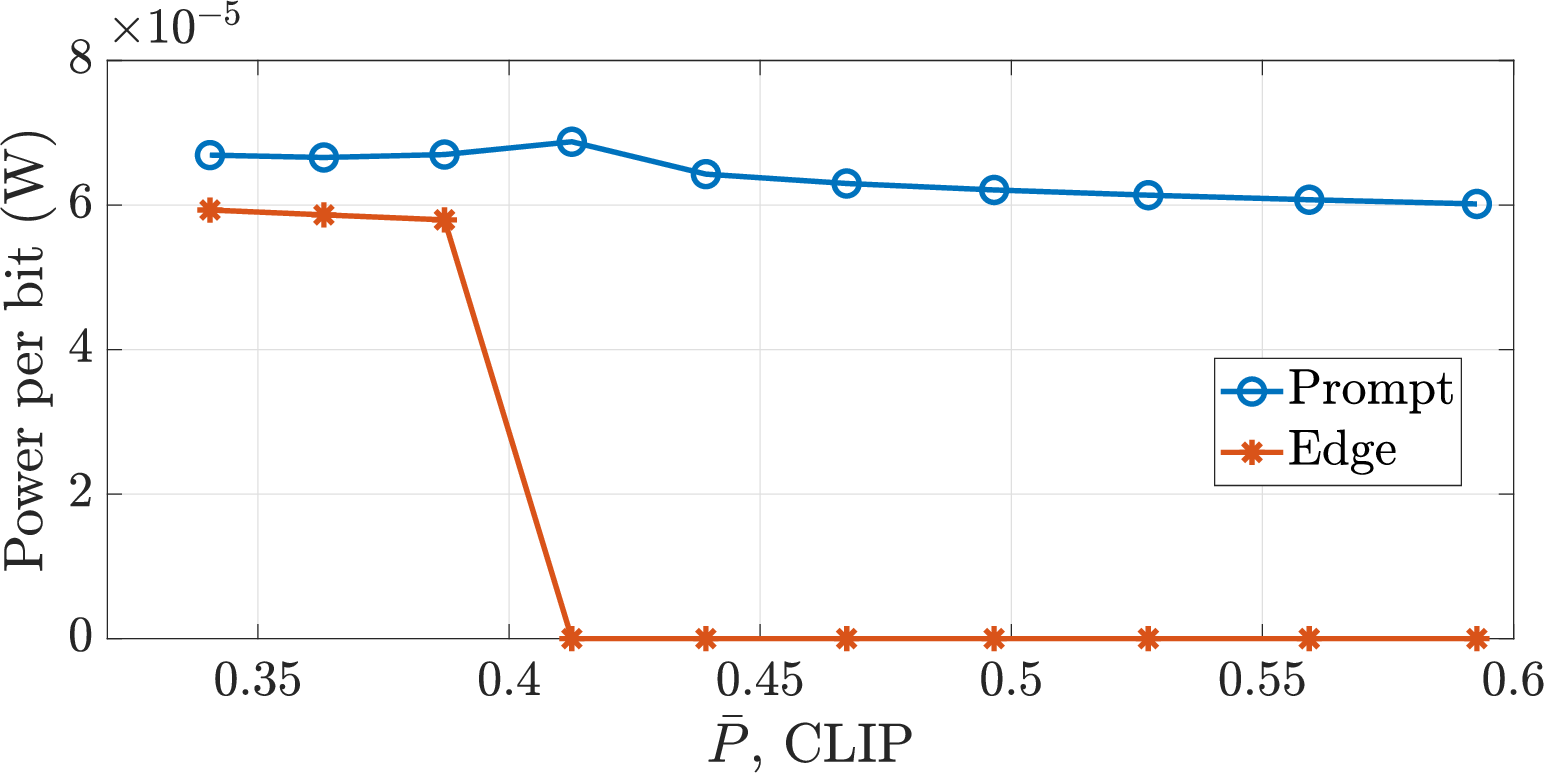}}
	%\subfigure[]{\includegraphics[width=0.45\columnwidth]{figures/power_bit_uncoded_forward_msssim.eps}}
	%\subfigure[]{\includegraphics[width=0.3\columnwidth]{figures/power_bit_uncoded_discard_msssim.eps}}
	\subfigure[]{
		\includegraphics[width=0.9\columnwidth]{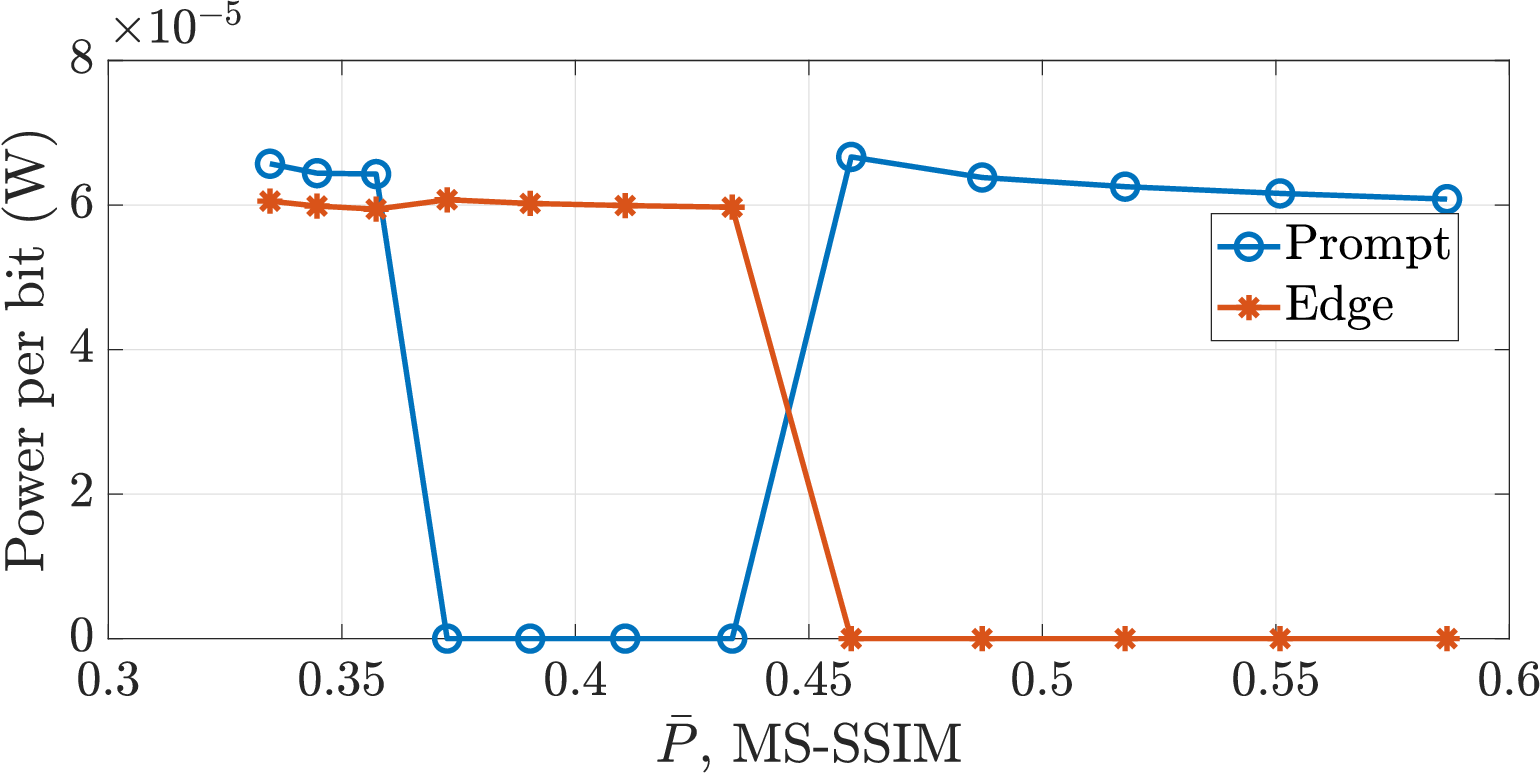}}
	\caption{{\color{sblue}Average power per bit: (a). Coded discard-with-error scheme with the CLIP metric. (b). Coded discard-with-error scheme with the MS-SSIM metric.}}
	\label{fig:power_bit_barP}
	
\end{figure}
\begin{figure*}[tbp]
	\centering
	\subfigure[]{
		\includegraphics[width=0.45\textwidth]{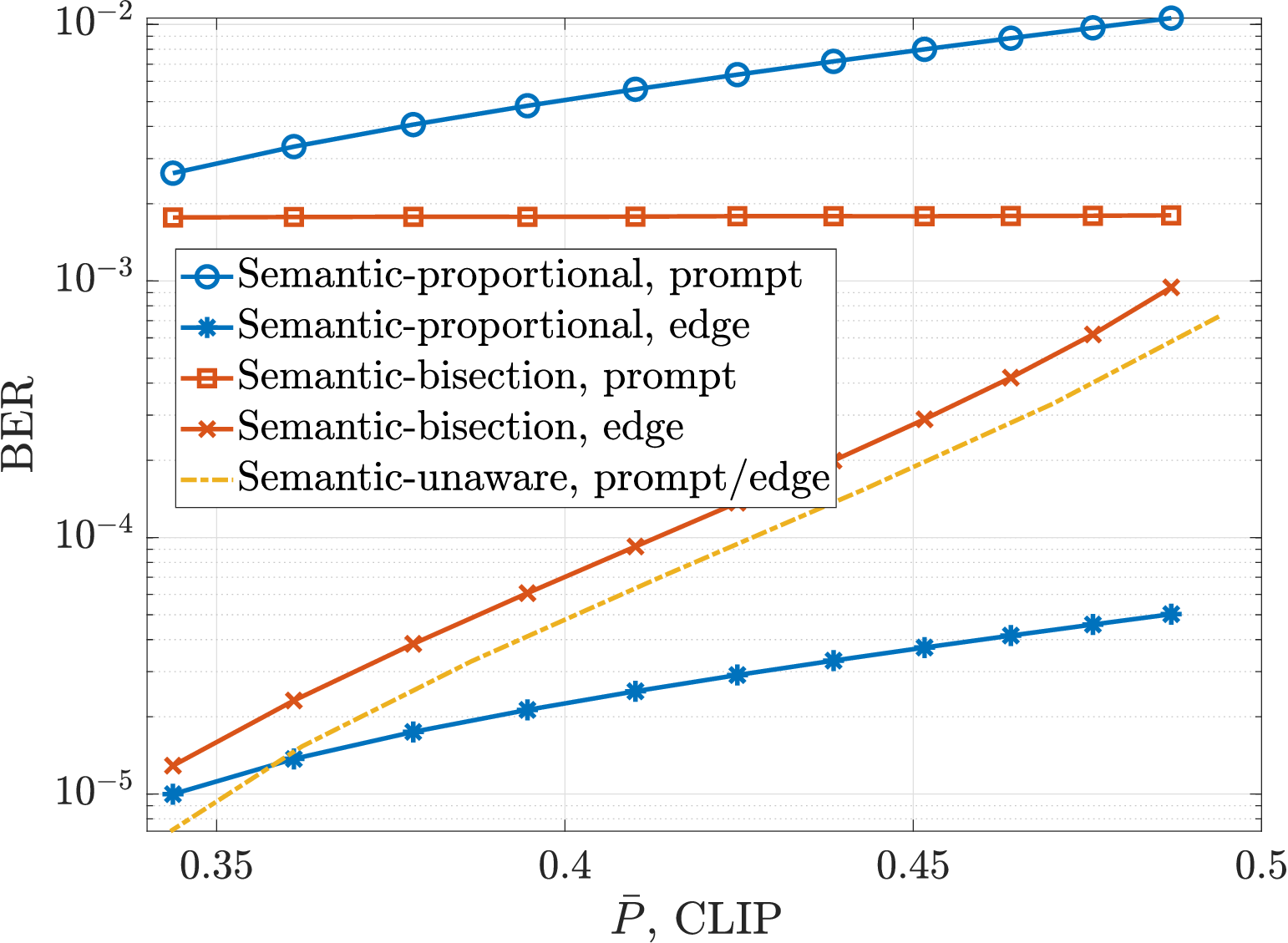}}
	\qquad
	\subfigure[]{\includegraphics[width=0.45\textwidth]{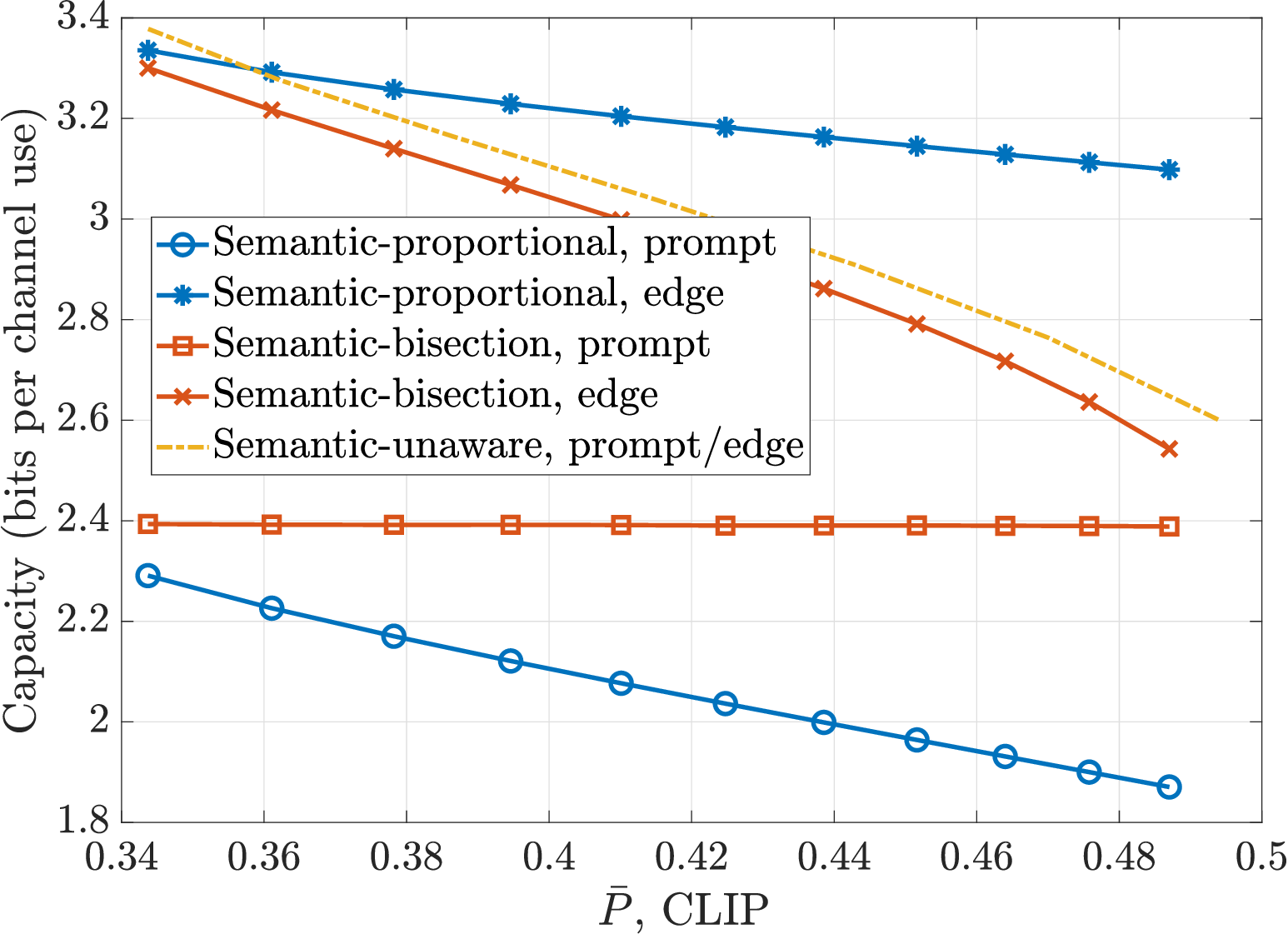}}
	\hspace{2em}
	\subfigure[]{
		\includegraphics[width=0.45\textwidth]{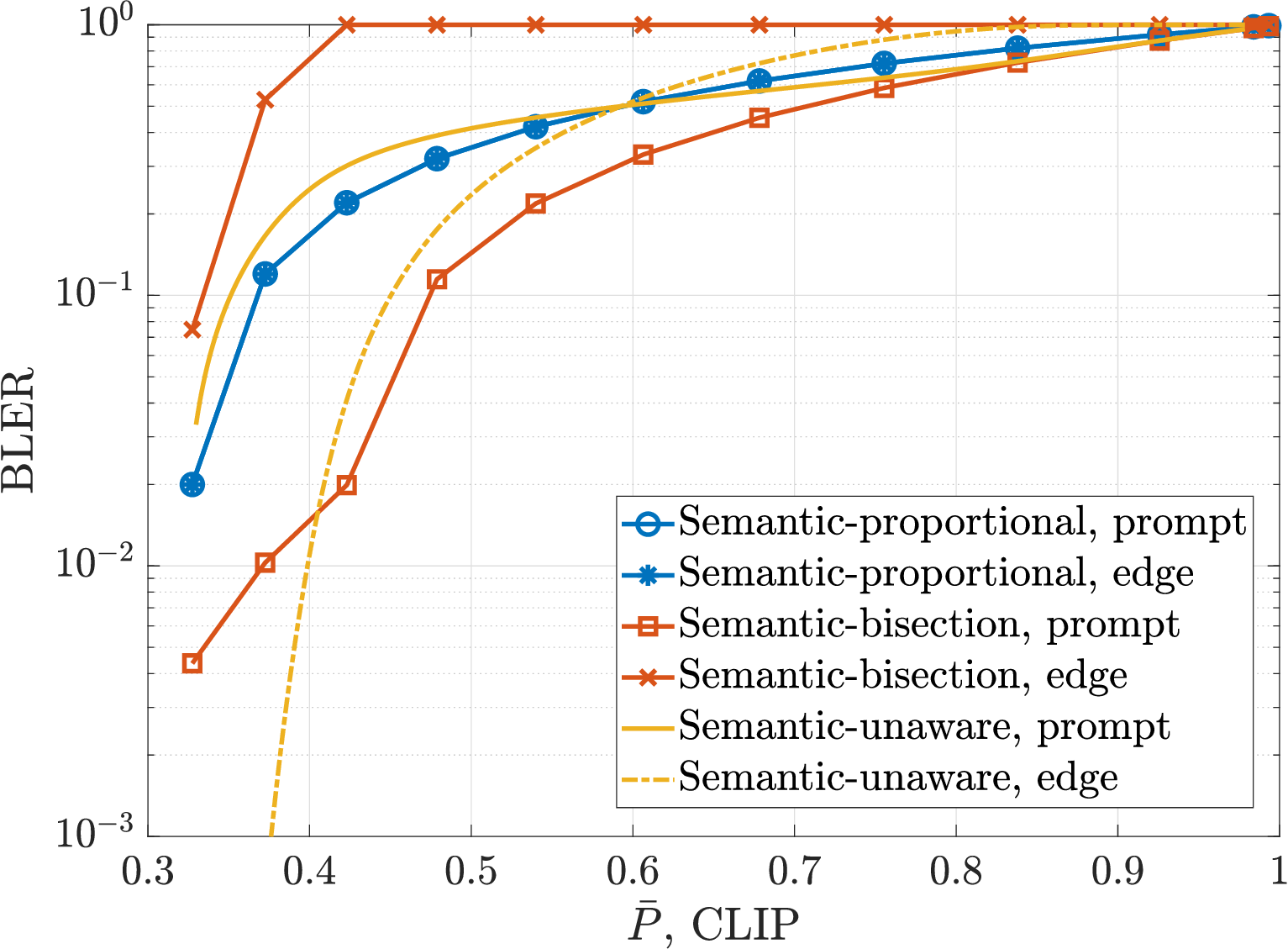}}
	\qquad
	\subfigure[]{\includegraphics[width=0.45\textwidth]{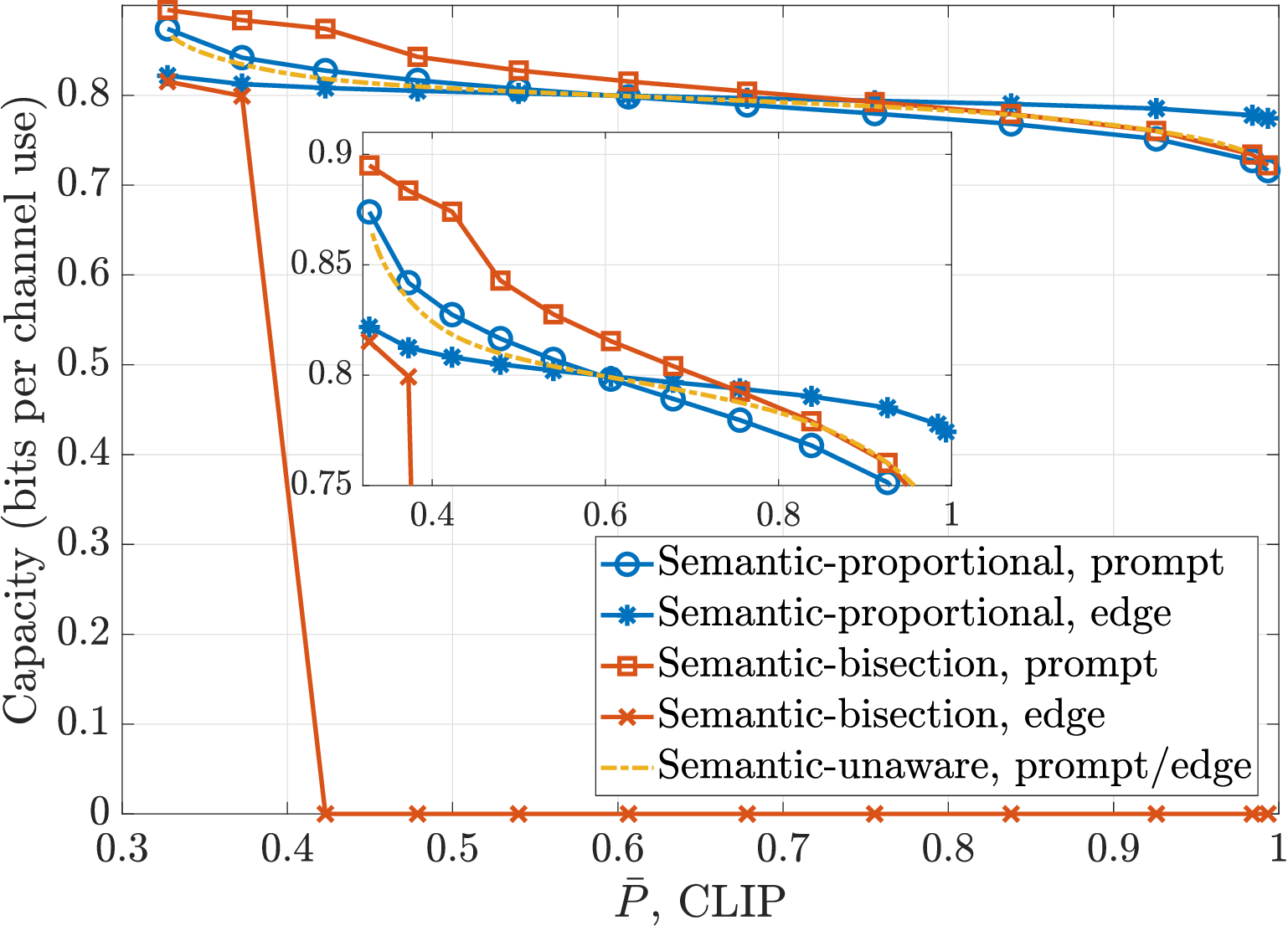}}
	\caption{{\color{sblue}Comparisons of transmission error and capacity: (a). BER under  uncoded forward-with-error scheme. (b). Capacity under uncoded forward-with-error scheme. (c). BLER under  coded discard-with-error scheme. (d). Capacity under discard-with-error chemes.}}
	\label{fig:ber_bler_capacity_barP}
\end{figure*} 

\subsection{\color{sblue}{Performance Assessment}}
To assess the proposed semantic-aware power allocation methods, we compare them against conventional semantic-unaware approaches {\color{sblue} in terms of the total power consumption, transmission error and channel capacity}. The proposed and traditional baseline methods are listed as follows:
\begin{itemize}
    \item Semantic-proportional: The allocated power is obtained based on Theorem \ref{lem5}, where semantic value requirements satisfy $\bar{L}_i/L_i=\bar{L}_j/L_j, \forall i,j\in \mathcal I$. 
    \item Semantic-bisection: The allocated power is obtained based on Algorithm \ref{alg1}.
    \item Semantic-unaware: {\color{sblue} The allocated power is obtained to ensure equal SNRs for all semantic data streams, resulting in identical BER and capacity performance under the channel-uncoded case, and identical capacity under the channel-coded case.}
\end{itemize}

{\color{sblue}Fig. \ref{fig:power_barP} compares total power consumption under specified perception performance requirements. The proposed semantic-bisection method achieves significant power savings compared to the baseline: up to 10\% in channel-uncoded cases and 90\% in channel-coded cases. Under stringent semantic requirements,  the semantic-proportional method achieves similar performance to the semantic-bisection method. However, its advantage over the semantic-unaware method diminishes as $\bar{P}$ increases. In the channel-coded case, the  semantic-bisection method shows sharp power reductions at a certain semantic requirement $\bar{P}$, corresponding to when power allocation to one semantic stream becomes zero as illustrated in Fig. \ref{fig:power_bit_barP}.}  While the edge map feature shows  higher semantic similarity under the MS-SSIM metric than the prompt, the prompt feature proves more valuable when considering semantic information per bit due to its shorter length. This leads to prioritizing  prompt transmission at high $\bar{P}$. However, in the range of $0.3313 \le \bar{P} \le 0.4720$, the prompt feature is not transmitted as it alone cannot meet such requirements.  

{ \color{sblue}Fig. \ref{fig:ber_bler_capacity_barP} compares transmission error and channel capacity performance under specified CLIP perception performance requirements. Under the semantic-unaware method, prompt and edge semantic data streams achieve identical capacities in both channel-uncoded and channel-coded cases due to equal received SNRs. While their BERs are identical in the channel-uncoded case,  BLERs differ in the channel-coded case due to their varying sequence lengths. For the channel-uncoded case,  the semantic-unaware method achieves lower BERs and higher capacities, as it requires higher power consumption to obtain the required semantic performance $\bar{P}$. 
Compared to the semantic-proportional method, it exhibits higher BER and lower capacity for edge feature transmission. 
In the channel-coded case, the semantic-proportional method achieves identical BLERs for both features due to the setting of $\bar{L}_i/L_i=\bar{L}_j/L_j, \forall i,j\in \mathcal I$. The semantic-unaware method shows lower BLER and higher capacity for edge feature transmission compared to semantic-bisection methods, with opposite results for the prompt feature. When $\bar{P}$ exceeds a threshold, 
edge map  transmission shows  BLER of $1$ and capacity of $0$, indicating no power allocation. These results suggest selective feature transmission based on channel conditions can optimize resource usage, potentially enabling adaptive semantic coding rates through selective semantic extractor activation. This is further supported by Fig. \ref{fig:cdf}, which shows the cumulative distribution function (CDF) of semantic performance across various power budgets in the channel-coded case.}

%To show how channel conditions impact the semantic performance, Fig. \ref{fig:cdf} gives the cumulative distribution function (CDF) of the achieved semantic performance under different total power budgets in terms of CLIP and MS-SSIM metrics. It shows that our proposed semantic-aware method significantly outperforms the conventional approach under the coded discard-with-error scheme. This initially demonstrates the potential of adaptive semantic coding rate to the channel conditions. 
%The advantage over the semantic-unaware method is evident in achieving semantic performance with higher perception, validating the effectiveness of adapting semantic encoding rate to the channel conditions.      

\begin{comment}
\begin{figure*}[tbp]
\centering
\subfigure[]{
\includegraphics[width=0.32\textwidth]{figures/power_ber_bpsk_forward_msssim.eps}}
\subfigure[]{
\includegraphics[width=0.32\textwidth]{figures/power_ber_bpsk_discard_msssim.eps}}
\subfigure[]{\includegraphics[width=0.32\textwidth]{figures/power_bler_coderate0.8_msssim.eps}}
\caption{Power consumption V.S. the perceptual performance $\bar P$ in terms of MS-SSIM metric. (a). Uncoded BPSK under forward-with-error scheme. (b). Uncoded BPSK under discard-with-error scheme. (c). Coded BPSK under discard-with-error scheme.}
\label{fig:power_barP_msssim}
\end{figure*}
\end{comment}

%\hre{To well demonstrate the generative performance for image transmission task, we present the generative image under different methods ensuring equal total power consumption. }

\begin{figure*}[tbp]
\centering
\subfigure[]{
\includegraphics[width=0.45\textwidth]{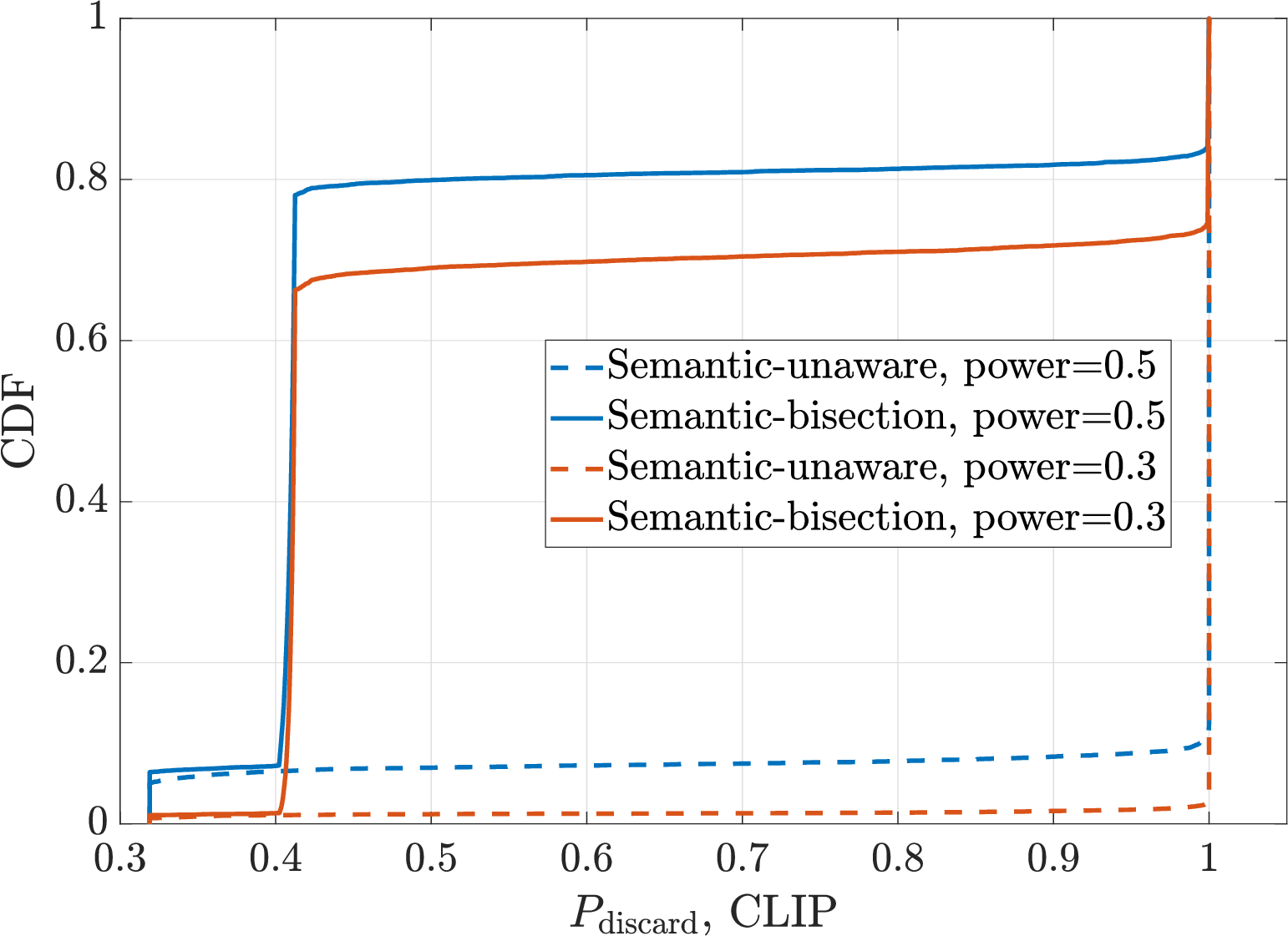}}
\qquad
\subfigure[]{
\includegraphics[width=0.45\textwidth]{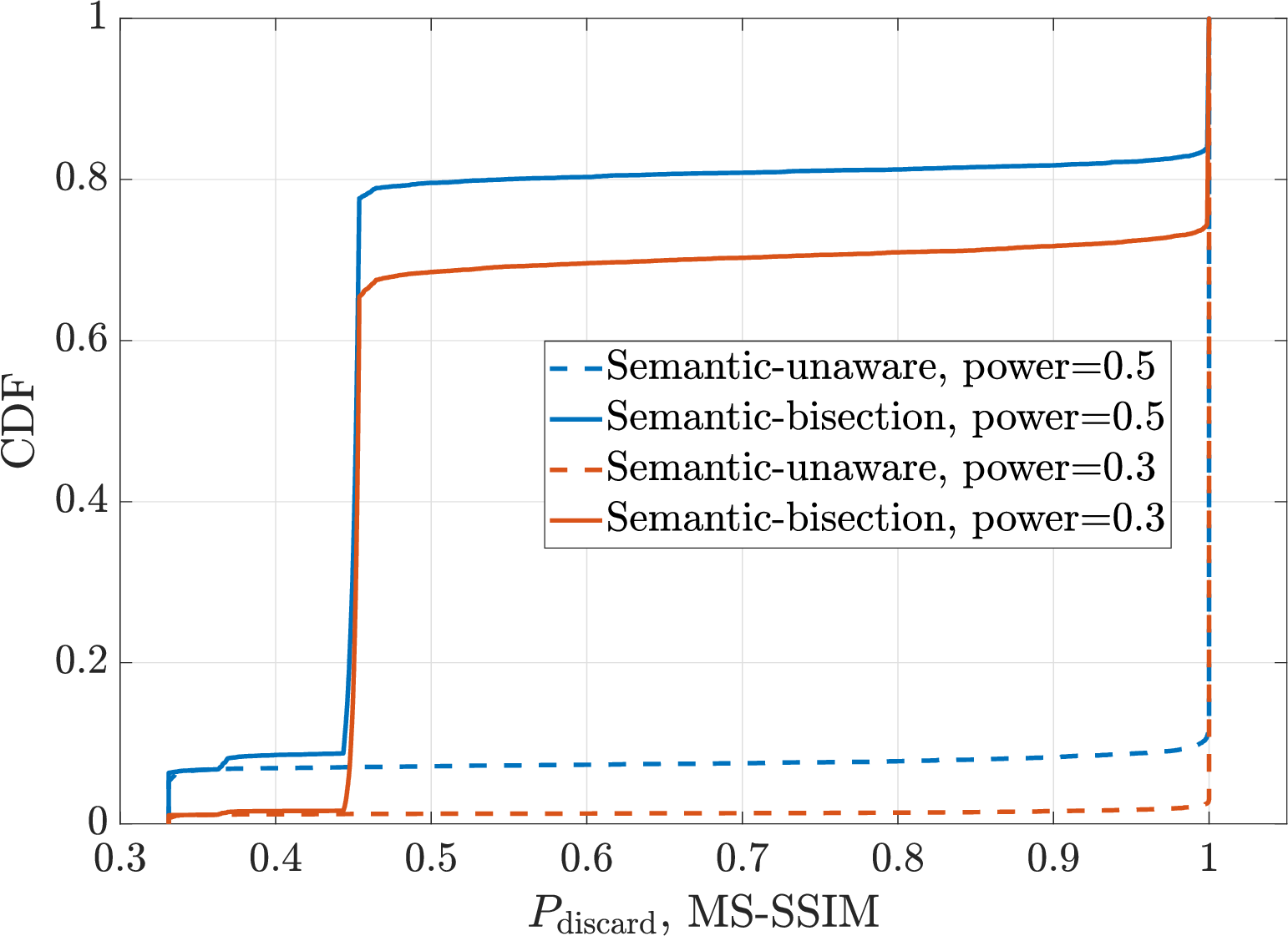}}
\caption{{\color{sblue}CDF of the achieved perception value under the coded discard-with-error scheme: (a). CLIP metric. (b). MS-SSIM metric.}}
\label{fig:cdf}
\end{figure*}

\section{Conclusion}

{\color{sblue} This paper presented a generative SemCom framework using pre-trained foundation models, where uncoded forward-with-error and coded discard-with-error schemes were proposed for the semantic decoder. The relationship between transmission reliability and regenerated signal quality was characterized through rate-distortion-perception theory, enabling semantic value quantification of semantic similarity between semantic data streams and the original source. Semantic-aware power allocation methods were developed for ultra-low rate and high fidelity SemComs to minimize power consumption while maintaining semantic performance. Simulations on the Kodak dataset validated the effectiveness of the proposed generative SemCom framework and verified the perception-error functions and semantic values. The proposed semantic-aware method was demonstrated to outperform the conventional approach, particularly in the channel-coded case. A key finding revealed that  power allocated to certain semantic data streams could be zero, suggesting potential computation savings through selective deactivation of semantic extractors. This insight points toward future research in adaptive semantic communications with channel feedback to improve both radio and computation efficiency. %However, combining link adaptation with generative SemCom presents non-trivial challenges, opening new research directions in this field.
}

\appendices
\section{Proof of Lemma \ref{lem2}}\label{Proof_Lemma2}
\begin{proof}
{\color{sblue}According to the chain rule of mutual information \cite{thomas2006elements}, $I(\mathbf K; \hat{\mathbf{K}})$ under uncoded forward-with-error scheme satisfies:	
\begin{align}
	I(\mathbf K_i; \hat{\mathbf{K}}_i)&=\sum_{j=1}^{K_i}I(\mathbf K_{ij}; \hat{\mathbf{K}}_i | \mathbf K_{i1},\dots, \mathbf K_{i(j-1)})\nonumber\\
	&=\sum_{j=1}^{K_i}I(\mathbf K_{ij}; \hat{\mathbf{K}}_i) 	 
\end{align}where the second equality holds due to bit independence as per Assumption \ref{assp2}. Further applying the chain rule to $I(\mathbf K_{ij}; \hat{\mathbf{K}}_i)$ yields:
\begin{align}
I(\mathbf K_{ij}; \hat{\mathbf{K}}_i)&= \sum_{j'=1}^{K_i}I(\mathbf K_{ij}; \hat{\mathbf{K}}_{ij'}|\hat{\mathbf{K}}_{i1},\dots, \hat{\mathbf{K}}_{i(j'-1)})\nonumber\\
	&=\sum_{j'=1}^{K_i}I(\mathbf K_{ij}; \hat{\mathbf{K}}_{ij'})\nonumber\\
	&= I(\mathbf K_{ij}; \hat{\mathbf{K}}_{ij}),
\end{align}where  the second equality follows from the interdependence among received bits within $\hat{\mathbf K}_i$, and the third equality holds because received bit $\hat{\mathbf K}_{ij'}$ is independent of $\hat{\mathbf K}_{ij}$ when $j'\neq j$. Given that $I(\mathbf K_{ij}; \hat{\mathbf{K}}_{ij})= H(\phi_{ij})-H(\psi_{ij})$  for a Bernoulli source transmitted over binary symmetric channels, we obtain:
\begin{align}
	I(\mathbf K_i; \hat{\mathbf{K}}_i) =\sum_{j=1}^{K_{i}}H(\phi_{ij})-H(\psi_{ij}),
\end{align}which decreases in BER $\psi_{ij}$ with $\psi_{ij}\le 0.5$.

For the coded discard-with-error scheme, since erroneous received data streams are discarded, we have $H(\mathbf K_i| \hat{\mathbf{K}}_i\neq\mathbf K_i)=H(\mathbf K_i)$. Consequently, the mutual information $I(\mathbf K; \hat{\mathbf{K}})$ can be expressed as:
\begin{align}
	I(\mathbf K_i; \hat{\mathbf{K}}_i)&=H(\mathbf K_i) - H(\mathbf K_i| \hat{\mathbf{K}}_i)\nonumber\\ &=H(\mathbf K_i)- \mathbb P(\hat{\mathbf{K}}_i=\mathbf K_i) H(\mathbf K_i| \hat{\mathbf{K}}_i=\mathbf K_i)\nonumber\\&\quad-P(\hat{\mathbf{K}}_i\neq\mathbf K_i) H(\mathbf K_i| \hat{\mathbf{K}}_i\neq\mathbf K_i)\nonumber\\
	&=H(\Phi_i)-\Psi_iH(\Phi_i),	 
\end{align}which decreases with the BLER $\Psi_i$. The proof is thus  completed.}      
\end{proof}

\section{Proof of Lemma \ref{lem5}}\label{Proof_Lemma5}
\begin{proof}
        The solutions to problems $\mathcal P1\text{-}1$ and $\mathcal P2\text{-}1$ can be readily obtained by substituting $\psi_{i}^*$ and $\Psi_{i}^*$ back to (\ref{eq:BER_p}) and (\ref{eq:BLER_polyanskiy}).   The difficulty in obtaining the optimal power allocation to problem $\mathcal P3\text{-}1$ lies in the transcendental equation of (\ref{eq:BLER_polyanskiy}). %such that
        %\begin{equation}
        %\ln2\sqrt{\frac{N_{i}}{V_{i}}}\left(C_{i}-\frac{K_{i}}{N_{i}}\right) = Q^{-1}\left(\Psi^*_{i}\left(p_{i}\right)\right),
        %\end{equation}
        Letting $\alpha_i\triangleq \frac{Q^{-1}\left(\Psi^*_{\mathbf K_i}\right)}{\sqrt{N_i}}$, (\ref{eq:BLER_polyanskiy}) can be rewritten as 
        \begin{equation}\label{eq:Lambert}
        \ln{\left((1+{\color{sblue}\mathrm{snr}_i})e^{-\frac{K_i}{N_i}}\right)}-\alpha_i\sqrt{1-(1+{\color{sblue}\mathrm{snr}_i})^{-2}}=0.
        \end{equation}Letting $\eta_i\triangleq \ln{\left((1+{\color{sblue}\mathrm{snr}_i})e^{-K_i/N_i}\right)}$ and $\beta_i=e^{-K_i/N_i}$, we have $1+\mathrm{SNR}_i=\beta_i^{-1}e^\eta_i$.  (\ref{eq:Lambert}) can then be further rewritten by
        \begin{equation}
       \eta_i=\alpha_i\sqrt{1-\beta_i^{2}e^{-2\eta_i}},
        \end{equation}which can be further expressed in a generalized Lambert W function fashion by
        \begin{equation}
       -4\beta_i^2\alpha_i^2= (2\eta_i-2\alpha_i)(2\eta_i+2\alpha_i)e^{2\eta_i}.
        \end{equation}The solution of $\eta_i^*$ is denoted as $\eta_i^* = W\left(^{2\alpha_i}, ^{-2\alpha_i} ;-4\beta_i^2\alpha_i^2\right)/2$. Thus, the optimal power $q_i^*$ is given by
       \begin{equation}
       q_i^*= \frac{\sigma_{i}^{2}}{2\vert h_{i}\vert^2}\left(e^{\frac{K_i}{N_i}+\eta_i^*}-1\right).
       \end{equation}
    \end{proof}

\section{Visual quality of the regenerated images}\label{generated_image}
Fig. \ref{fig:examples} depicts the regenerated images of Kodim01 and Kodim21 with transmission errors. The left three columns are the regenerated images under the uncoded forward-with-error scheme, while the rightmost column shows the regenerated images under the coded discard-with-error scheme.  %The compression rates featured by bit per pixel (BPP) under the proposed generative SemCom are $0.0278$ and $0.0260$, respectively. This indicates that ultra-low rates can be achieved by the proposed generative SemCom framework.  The leftmost column depicts the source images, and their textual prompt and edge map semantic features. 
%The left three columns are the regenerated images under the uncoded forward-with-error scheme, showing that the visual quality is degrading with the error. The rightmost column shows the regenerated images under the coded discard-with-error scheme, demonstrating that the textual prompt better provides the semantic contents while the edge map captures the structural similarity.
\begin{figure*}[th]
\centering
\subfigure[Kodim01]{
\includegraphics[width=0.485\textwidth]{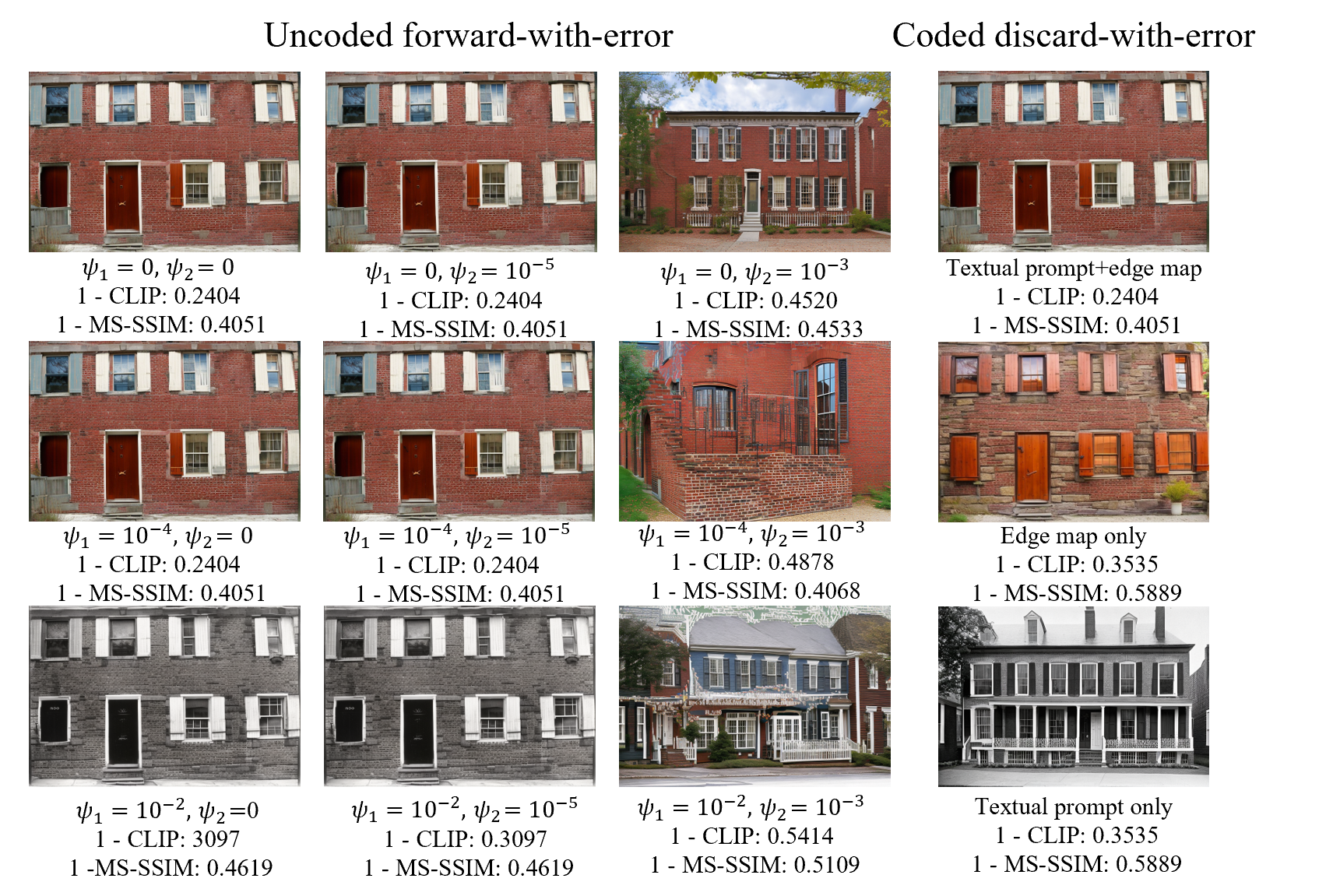}
}
\subfigure[Kodim21]{
\includegraphics[width=0.485\textwidth]{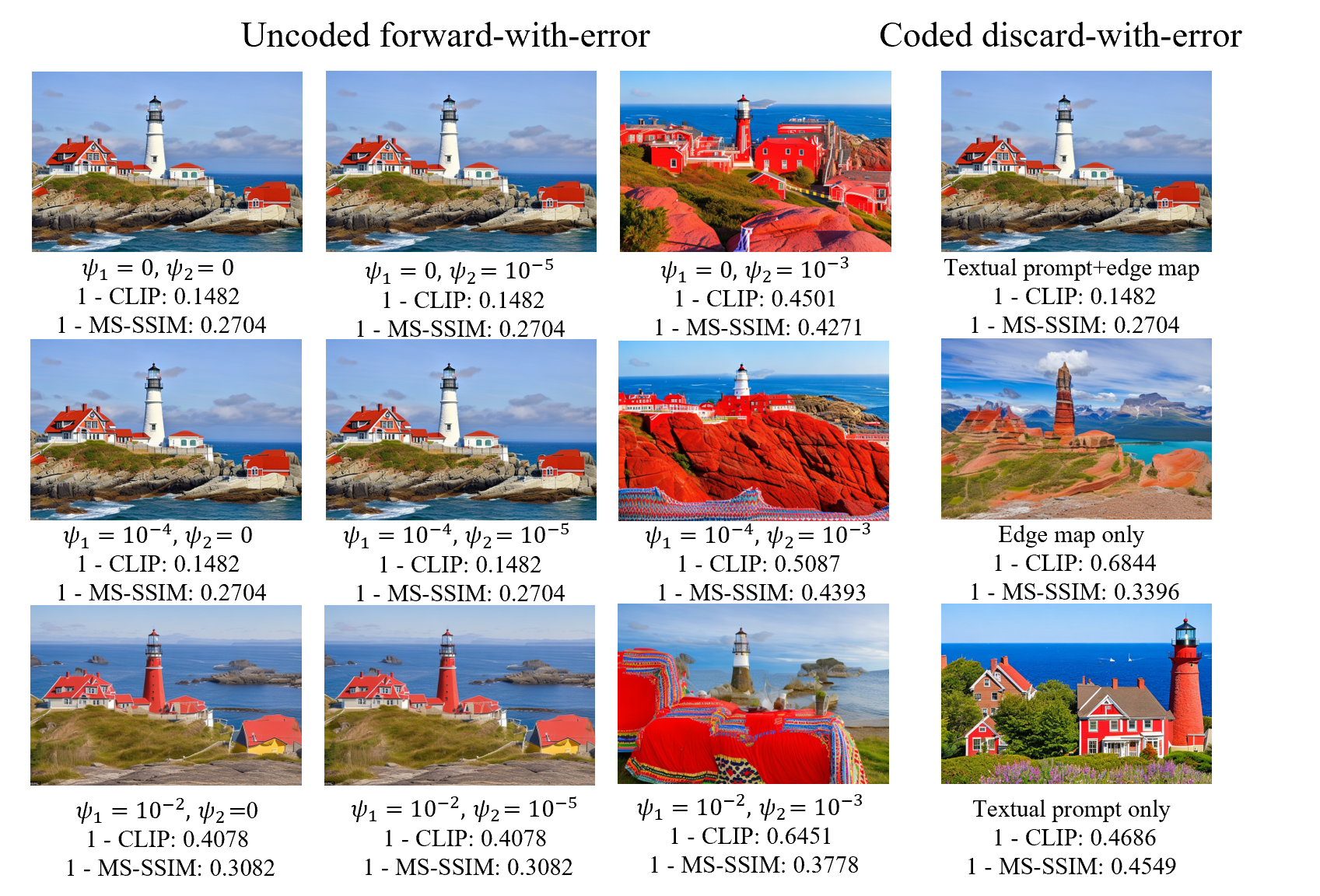}}
\caption{Regenerated images under uncoded forward-with-error and coded discard-with-error schemes: (a). Kodim01. (b). Kodim21.} %\hre{Use Kodim 21 to make comparisons with JPEG, JPEG2000, Cheng2020, Mbt2018  in terms of the BPP and the CLIP value} }
\label{fig:examples}
\end{figure*}

\begin{comment}
\section*{Acknowledgment}
{\color{sblue}This work was supported by the U.K. Department for Science, Innovation, and Technology under Project TUDOR (Towards Ubiquitous 3D Open Resilient Network).  The authors would like to thank Jinfei Wang and Li Qiao for their assistance with the simulation environment.}
\end{comment}
\bibliographystyle{IEEEtran}
\bibliography{semantic}

\end{document}